\documentclass{jfm}
\usepackage{graphicx}
\usepackage{newtxtext}
\usepackage{newtxmath}
\usepackage{natbib}
\usepackage{hyperref}
\usepackage{xcolor}
\usepackage{cancel}
\usepackage{pifont}
\usepackage{tikz}
\usetikzlibrary{tikzmark}
\tikzset{mycircled/.style={circle,draw,inner sep=0.1em,line width=0.1em}}
\hypersetup{
    colorlinks = true,
    urlcolor   = blue,
    citecolor  = blue,
}

\newcommand{\RomanNumeralCaps}[1]

\shorttitle{Transient rod-climbing in an Oldroyd-B fluid}
\shortauthor{T. Ruangkriengsin, R. Brandão, K. Wu, J. Hwang, E. Boyko, and H.A. Stone}

\title{\Large Transient rod-climbing in an Oldroyd-B fluid}
 \author{Tachin Ruangkriengsin\aff{1},
Rodolfo Brandão\aff{2},
Katie Wu\aff{3},
Jonghyun Hwang\aff{3},
Evgeniy Boyko\aff{4} \and 
Howard A. Stone\aff{3}
 \corresp{\email\href{mailto:hastone@princeton.edu}{hastone@princeton.edu}}
}
  \affiliation{\aff{1}Program in Applied and Computational Mathematics, Princeton University, Princeton, NJ 08544, USA
  \aff{2}School of Mathematics, University of Bristol, Fry Building, Woodland Road, Bristol BS8 1UG, UK
  \aff{3}Department of Mechanical and Aerospace Engineering, Princeton University, Princeton, NJ 08544, USA
  \aff{4}Faculty of Mechanical Engineering, Technion – Israel Institute of Technology, Haifa 3200003, Israel}
  
\begin{document}
\maketitle

\begin{abstract}
The Weissenberg effect, or rod-climbing phenomenon, occurs in non-Newtonian fluids where the fluid interface ascends along a rotating rod. Despite its prominence, theoretical insights into this phenomenon remain limited. In earlier work, Joseph \& Fosdick (\emph{Arch. Rat. Mech. Anal.}, vol. 49, 1973, pp. 321--380) employed domain perturbation methods for second-order fluids to determine the equilibrium interface height by expanding solutions based on the rotation speed. In this work, we investigate the time-dependent interface height through asymptotic analysis with dimensionless variables and equations using the Oldroyd-B model. We begin by neglecting surface tension and inertia to focus on the interaction between gravity and viscoelasticity. In the small-deformation scenario, the governing equations indicate the presence of a boundary layer in time, where the interface rises rapidly over a short time scale before gradually approaching a steady state. By employing a stretched time variable, we derive the transient velocity field and corresponding interface profile on this short time scale and recover the steady-state profile on a longer time scale. Subsequently, we reintroduce small but finite inertial effects to investigate their interplay with viscoelasticity and propose a criterion for determining the conditions under which rod-climbing occurs.
\end{abstract}


\section{Introduction}

The rod-climbing phenomenon, also known as the Weissenberg effect \citep{weissenberg}, is one of the most iconic examples of non-Newtonian fluid behavior in which the fluid's surface climbs along a rotating rod. Heuristically, this effect arises from the interplay between elastic and viscous forces in complex fluids, where the unequal normal stresses within the fluid generate a hoop stress around the rod, driving an upward motion~\citep{bird1987dynamics1}. Its visually compelling nature has made the rod-climbing phenomenon a foundational concept in the study of non-Newtonian fluids, serving as an essential tool in both pedagogical and research contexts~\citep{ewoldt2022designing}. Although widely recognized, theoretical insights into this phenomenon remain limited, owing to the challenges in modeling the constitutive equations for such fluids.

The earliest theoretical attempt to explain the rod-climbing phenomenon dates back to the work of \citet{serrin1959}. Serrin derived the unidirectional Couette flow solution for second-order fluids. By neglecting the non-equilibrated shear stress at the free surface, Serrin calculated the rise height and identified the conditions under which the fluid's free surface would climb the rod. A similar calculation, introduced by \citet{Giesukus1961} two years after Serrin's work, neglects inertial effects but applies to a wider range of constitutive equations; however, the issue of non-equilibrated shear stress along the free surface remains unresolved.  Both Serrin and Giesekus  asserted that their solutions would remain valid, approximately, as long as the free surface remained relatively horizontal. Building on this concept of small deformations, \citet{joseph1973} developed a systematic framework for constructing the steady free-surface shape of second-order fluids using the domain perturbation method. This solution involved a perturbation series for both the flow profile and the fluid domain based on a prescribed (small) angular velocity $\Omega$. In subsequent works,~\citet{joseph1973partII} and \citet{beaver1980} compared their theoretical predictions with the experimental observations, finding good agreement. In fact, the theoretical work of \citet{joseph1973}, completed over 50 years ago, continues to be widely used for rheological measurements today \citep{more2023rod}. Building on the pioneering work of \citet{joseph1973}, several studies have extended the original framework using perturbative approaches. For example, \citet{Yoo_Joseph_Beavers_1979} developed a higher-order theory of the rod-climbing phenomenon by expanding the perturbation series to $O(\Omega^4)$. Meanwhile, \citet{SIGINER1984} analyzed the free surface of second-order fluids between vertical cylinders rotating about non-concentric axes, utilizing the domain-perturbation method and bipolar coordinates to aid the analysis.

Beyond its theoretical significance, the rod-climbing phenomenon has substantial practical applications in rheology. In particular, several studies highlighted its utility in rheometric applications to determine key material parameters, such as normal stress coefficients and relaxation times, by analyzing climbing heights and rates~\citep{Beavers_Joseph_1975, Choi1991Relaxation, Choi1992Relaxation}. More recently, \citet{more2023rod} revisited rod-climbing rheometry with the aid of a modern torsional rheometer. By integrating rod-climbing experiments with small-amplitude oscillatory shear flow measurements and steady-shear measurements from commercial rheometers, they successfully predicted the normal stress coefficients of a polymer solution at low shear rates, extending below the sensitivity range of conventional rheometers.

To date, all theoretical studies on the rod-climbing phenomenon have focused exclusively on its steady-state behavior and have been limited to second-order fluid models. However, the use of second-order fluids comes with several limitations. First, the retarded-motion expansion (e.g., second-order and third-order fluid models) is only valid for small Deborah or Weissenberg numbers~\citep{bird1987dynamics1}; applying it beyond this range can lead to unphysical results. Second, second-order fluid models are highly unstable. For instance, a steady shear flow in a second-order fluid is unstable to short-wavelength perturbations, which, when combined with time-dependent flows, lead to local exponential growth in stress~\citep{Intro_C_F}. In this work, we focus on investigating the transient dynamics of the rod-climbing phenomenon. Access to a time-dependent profile enables a more accurate application of rod-climbing rheometry, providing a valuable tool for both advancing the theoretical understanding of the flows of non-Newtonian fluids and leveraging the results to determine material properties. As noted above, to avoid the issues of using the second-order fluids model with an unsteady flow, we opt to use the Oldroyd-B equation~\citep{oldroyd1950formulation, bird1987dynamics1}.  

The Oldroyd-B constitutive equation is used widely for characterizing the behavior of constant shear-viscosity viscoelastic fluids, such as Boger fluids~\citep{james2009boger}. In 1950, Oldroyd calculated the forces across a horizontal plane needed to maintain steady two-dimensional flow around a vertically oriented rotating rod, using his recently derived constitutive equations~\citep{oldroyd1950formulation, oldroyd1951}. He discovered that the Oldroyd-B model, which is now known to apply approximately to polymer solutions, would result in fluid climbing the rod, while the Oldroyd-A model, which involves a sign change~\citep{hinchharlen}, would cause the fluid to descend~\citep{oldroyd1950formulation}. Beyond the physical considerations, it is important to note that the Oldroyd-B model can be derived from first principles by analyzing the stress in a dilute polymer solution, where the polymer is represented as a dumbbell structure of two beads connected by a linear elastic spring. Such representation offers valuable physical intuition for understanding the viscoelastic effects arising from the interaction between fluid flow and the stretching or compressing of suspended polymers, the former which is known to give rise to hoop stresses in flows with curved streamlines. 

There have been a few theoretical studies on the start-up flow of an Oldroyd-B fluid between two rotating cylinders. These studies will be particularly relevant later in our paper when discussing unsteady flow. For example, \citet{fetecau2005} uses the Fourier--Bessel series to derive the Couette flow profile, assuming the flow depends only on the radial direction. However, similar to the steady flow calculations for two rotating cylinders in Oldroyd-B fluids by \cite{oldroyd1951}, no consideration is given to the free-surface interface at the top of the fluid. To conclude this section, we summarize in table~\ref{reftable} the previous theoretical works on flows near a rotating rod in complex fluids, with and without accounting for the free surface.
\begin{table}
\begin{center}
\renewcommand{\arraystretch}{1.2}
\begin{tabular}{lccccp{0.17\linewidth} p{0.21\linewidth}}
Model                                  & \multicolumn{2}{c}{Flow field} & \multicolumn{2}{c}{Rise height} & Remarks                             & Reference          \\
\multicolumn{1}{c}{}                   & Steady        & Unsteady       & Steady        & Unsteady        &                                     &                    \\
                                       &               &                &               &                 &                                     &                    \\
\multicolumn{1}{c}{Second-order fluid} & Yes           & No             & Yes*          & No              & *partial \mbox{calculations}               & \mbox{\citet{serrin1959}}, \mbox{\citet{Giesukus1961}}   \\
                                       & Yes           & No             & Yes           & No              & \mbox{include inertia and} \mbox{surface tension} & \mbox{\citet{joseph1973}} \\
                                       & Yes           & No             & Yes           & No              & \mbox{higher-order} \mbox{approximation}          & \citet{Yoo_Joseph_Beavers_1979}                \\
                                       & Yes           & No             & Yes           & No              & \mbox{non-concentric} \mbox{cylinders}            & \citet{SIGINER1984}            \\
                                       &               &                &               &                 &                                     &                    \\
Oldroyd-B                              & Yes           & No             & No            & No              & \mbox{obtained vertical} \mbox{normal stresses}   & \mbox{\citet{oldroyd1950formulation}}, \mbox{\citet{oldroyd1951}}            \\
                                       & Yes           & Yes            & No            & No              & start-up flow                       & \citet{fetecau2005}            \\
                                       & Yes           & Yes            & Yes           & Yes             &                                     & Current Work      
\end{tabular}
  \caption{Theoretical studies on flows around a vertically oriented rotating rod in complex fluids. The work of \citet{joseph1973} has been reproduced using modern notation by \citet{more2023rod}.}
  \label{reftable}
\end{center}
\end{table}

The remaining sections of this paper are organized as follows: In $\mathsection$~\ref{formulation}, we present the problem formulation, including the scalings for the variables and the governing equations. In $\mathsection$ \ref{setup}, we outline the preliminary setup and introduce the techniques used in later sections, such as the distinguished limit of the problem, the perturbation expansion in Weissenberg number, and the domain perturbation method for simplifying the interface boundary conditions. In $\mathsection$~\ref{steady_section}, we exclude inertial effects to focus solely on the interaction between viscoelasticity and gravity, deriving the steady interface profile that represents the long-time free-surface behavior. In $\mathsection$ \ref{transient_section}, we extend the steady interface analysis by examining the short-time scale to obtain the transient interface profile. In $\mathsection$~\ref{with_inertia_section}, we reintroduce small inertial effects to explore the interplay between inertia and viscoelasticity, and examine the conditions under which the fluid climbs the rod. We conclude with a discussion of the results in $\mathsection$ \ref{conclusions}.

\section{Problem formulation}\label{formulation}

\begin{figure}
\centering
\includegraphics[scale=1.4]{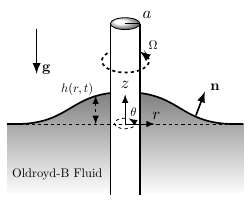}
\caption{Schematic illustration of an infinite rod with radius $a$ rotating with an angular speed $\Omega$ in a viscoelastic fluid.}
\label{problem_schematic}
\end{figure}

We investigate the rotation of an infinitely long rod with angular velocity $\Omega$, where the rod has radius $a$ and its axis is aligned along the $z-$direction, while immersed in a dilute viscoelastic polymer solution with density $\rho$. We assume that the flow is axisymmetric and employ 
cylindrical coordinates $(r,\theta,z)$ to examine the time-dependent free-surface profile  $h(r,t)$ of a viscoelastic fluid, with the reference height taken to be zero as $r\longrightarrow \infty$, as shown in figure~\ref{problem_schematic}. 
Our analysis primarily focuses on the case where the interface deformation is considered ``small," with the validity of this assumption to be clarified below through asymptotic analysis.

To model the viscoelastic contributions to the rod-climbing effect, we employ the Oldroyd-B constitutive model \citep{oldroyd1950formulation, bird1987dynamics1}.
We define a 
conformation tensor $\textbf{A}(r,z,t)$, representing the polymer's state of deformation,
where $\textbf{A} = \textbf{I}$ indicates the undistorted equilibrium state. As the polymer is suspended in the fluid, it experiences stretching, reorientation, and advection, which, in the Oldroyd-B model results in the evolution of the conformation tensor according to
\begin{equation}\label{evolution_conformation}
    \frac{\partial \textbf{A}}{\partial t} + \textbf{u} \cdot \nabla \textbf{A} - \textbf{A} \cdot \nabla \textbf{u} - (\nabla \textbf{u})^{\rm T}\cdot \textbf{A} = -\frac{1}{\lambda} (\textbf{A}-\textbf{I}),
\end{equation}
where $\textbf{u}$ is the velocity vector and $\lambda$ is the relaxation time of the polymer. The first two terms on the left-hand side of (\ref{evolution_conformation}) represent the material derivative of the conformation tensor, capturing the advection of the polymer by the background flow. The next two terms describe the stretching of the polymer due to the local velocity gradient. On the other hand, the right-hand side of (\ref{evolution_conformation}) accounts for the polymer's relaxation back to its equilibrium state, with the rate determined by the relaxation time. When the polymer stretches beyond its equilibrium state, it generates polymeric stresses due to the tension in the spring. In the Oldroyd-B model, the total stress $\boldsymbol{\sigma}$ in the fluid can be expressed as
\begin{equation}\label{stress}
    \boldsymbol{\sigma} = -p \textbf{I} + 2\mu_s \textbf{E} + G(\textbf{A}-\textbf{I}),
\end{equation}
where the first term in (\ref{stress}) represents the pressure component, the second term in (\ref{stress}) denotes the viscous stress from the solvent with viscosity $\mu_s$, where $\textbf{E} = \frac{1}{2}\left(\nabla {\textbf{u}} + (\nabla \textbf{u})^{\rm T}\right)$ is the rate-of-strain tensor, and the last term in (\ref{stress}) accounts for the polymeric contribution to the stress with $G$ representing the shear modulus of the fluid. It is helpful to define $\mu_p = G\lambda$ as the polymer viscosity at zero shear rate; throughout this paper, we will use $G = \mu_p/\lambda$. 

In $\mathsection$$\mathsection$ \ref{steady_section} and  \ref{transient_section}, we shall assume that the Reynolds number defined by 
\begin{equation}\label{reynolds }
Re = \frac{\rho a^2 \Omega}{\mu_s + \mu_p},
\end{equation}
is small, i.e. $Re\ll 1$. This assumption is valid, for instance, when the rod's angular rotation speed is low, in line with the assumption that the deformation along the free-surface interface is small. Neglecting inertial effects, the continuity and momentum equations become
\begin{equation}\label{continuity_momentum}
    \nabla \cdot \textbf{u} = 0\quad\hbox{and} \quad \nabla \cdot \boldsymbol{\sigma} =  \rho g \textbf{e}_z,
\end{equation} where $\textbf{e}_z$ is directed vertically upwards.  
We will revisit the role of small but finite inertial effects in $\mathsection$~\ref{with_inertia_section}. Equations (\ref{evolution_conformation}), (\ref{stress}), and (\ref{continuity_momentum}) form the basis of the Oldroyd-B constitutive description and model flow equations that we investigate in the upcoming sections. These equations are supplemented with boundary and initial conditions 
applied to the rotating rod, the free surface, and the far-field region. First, we assume the no-slip boundary condition along the surface of the rotating rod,
\begin{equation}\label{bc_dim_rod}
    \textbf{u} = a\Omega\textbf{e}_{\theta} \quad \text{at} \quad r = a.
\end{equation}
In the far-field, $r \longrightarrow \infty$, we assume that the interface remains flat, the fluid velocity diminishes to zero, and the microstructure remains in its equilibrium state,
\begin{equation}\label{bc_dim_farfield}
h \longrightarrow 0, \quad \textbf{u} \longrightarrow \textbf{0}, \quad \textbf{A} \longrightarrow \textbf{I} \quad \text{as} \quad r \longrightarrow \infty.
\end{equation}
Next, we impose both the stress and kinematic boundary conditions along the free surface, $z = h(r,t)$, which has unit outward normal $\textbf{n}$. To better explore the dynamics of the rising interface, we focus solely on the interaction between gravity and viscoelasticity (and later allow for small inertia), neglecting surface tension, which allows us to disregard the capillary-static rise and the contact angle between the fluid and the rod. As such, the 
stress boundary condition along the interface, referencing the fluid pressure to the ambient pressure, becomes
\begin{equation}\label{bc_dim_stress}
    \textbf{n} \cdot \boldsymbol{\sigma} = \textbf{0} \quad \text{at} \quad z = h(r,t),
\end{equation}
whereas the kinematic boundary condition along the interface is given by
\begin{equation}\label{bc_dim_kinematic}
    \frac{\partial h}{\partial t} = u_z - u_r\frac{\partial h}{\partial r} \quad \text{at} \quad z = h(r,t).
\end{equation}
Finally, we assume that the interface is flat and the polymer is in its equilibrium state at the initial time,
\begin{equation}\label{bc_initial}
h = 0, \quad \textbf{A} = \textbf{I}  \quad \text{at} \quad t = 0.
\end{equation}
It is important to note that by making the low-Reynolds-number assumption we have neglected the time derivative of the velocity field, which prevents us from specifying the velocity at the initial time. 

\subsection{Scalings and governing equations in dimensionless form}\label{scaling_section}
Up to this point, we have presented the constitutive equations and their corresponding boundary conditions using dimensional variables. 
We now rescale all relevant variables. Specifically, we rescale lengths by $a$, velocities by $a\Omega$, stresses by $(\mu_s + \mu_p)\Omega$, and time by $1/\Omega$, 
\begin{alignat}{5}\label{scaling_length_velocity}
Z  & = \frac{z}{a},  &\qquad  R &  =  \frac{r}{a},  &\qquad  H & = \frac{h}{a}, &\qquad \textbf{U}  &=  \frac{\textbf{u}}{a\Omega},  \notag
\\
T &= t\Omega, &\qquad \boldsymbol{\mathit{E}} & = \frac{\textbf{E}}{\Omega}, &\qquad P & =\frac{p}{(\mu_s+\mu_p)\Omega}, &\qquad \boldsymbol{\Sigma}  &=  \frac{\boldsymbol{\sigma}}{(\mu_s+\mu_p)\Omega}.
\end{alignat}  
These dimensionless variables are accompanied by dimensionless parameters for viscosity ratios
\begin{equation}\label{viscosity_ratio}
\beta_p = \frac{\mu_p}{\mu_s + \mu_p} \quad \text{and} \quad \beta_s = 1 - \beta_p,
\end{equation}
the Weissenberg number, $Wi$, and the gravity parameter, $\mathcal{G}$,
\begin{equation}\label{wi_gravity_number}
Wi = \lambda \Omega \quad \text{and} \quad \mathcal{G} = \frac{\rho g a}{(\mu_s + \mu_p)\Omega}.
\end{equation}
The Weissenberg number, $Wi$, defined as the product of the fluid’s relaxation time and the characteristic shear rate, is commonly used to assess the viscoelastic properties of a material, indicating whether its behavior is more elastic or more viscous.  Another key parameter 
is the Deborah number, $De$, which represents the ratio of the material's response time to the characteristic time of the flow. In this problem, $Wi$ and $De$ are equivalent, so no distinction is made between the two. Here we focus on the case where 
$Wi \ll 1$, corresponding to the weak viscoelastic effects. The range of $\mathcal{G}$ will depend on the value of $Wi$, so that we can derive analytical expressions for the small deformation of the interface shape $H$, as will be discussed in $\mathsection$~\ref{distinguish}.

Using the scalings in (\ref{scaling_length_velocity}), the governing equations (\ref{evolution_conformation}), (\ref{stress}), and (\ref{continuity_momentum})  become
\begin{subequations}\label{dimensionless_eq}
\begin{align}
\nabla \cdot \textbf{U} = 0, \quad &\nabla \cdot \boldsymbol{\Sigma} = \mathcal{G} \textbf{e}_z,\label{continuity_momentum_dimensionless} \\ 
\boldsymbol{\Sigma} = -P \textbf{I} + 2\beta_s \boldsymbol{\mathit{E}} &+ \frac{\beta_p}{Wi}(\textbf{A}-\textbf{I}), \label{stress_dimensionless}\\
\frac{\partial \textbf{A}}{\partial T} +\textbf{U} \cdot \nabla \textbf{A} - \textbf{A} \cdot \nabla \textbf{U} - (\nabla \textbf{U})^{\rm T}\cdot \textbf{A} &= -\frac{1}{Wi} \left(\textbf{A}-\textbf{I}\right). \label{conformation_dimensionless}
\end{align}
\end{subequations}
Similarly, the corresponding boundary and initial conditions (\ref{bc_dim_rod})--(\ref{bc_initial}) become
\begin{subequations}\label{dimensionless_bc}
\begin{align}
\text{No-slip boundary condition:}\quad \textbf{U} = \textbf{e}_{\theta} \quad &\text{at} \quad R = 1,\label{bc_noslip_dimensionless}\\
\text{Normal-stress boundary condition:} \quad\textbf{n} \cdot \boldsymbol{\Sigma} = \textbf{0} \quad &\text{at} \quad Z = H(R,T), \label{normal_stress_bc} \\
\text{Kinematic boundary condition:} \quad \frac{\partial H}{\partial T} = U_z - U_r\frac{\partial H}{\partial R} \quad &\text{at} \quad Z = H(R,T),\label{kinematic_bc}\\
\text{Far field:} \quad  H \longrightarrow 0, \quad \textbf{U} \longrightarrow \textbf{0}, \quad \textbf{A} \longrightarrow \textbf{I} \quad &\text{as} \quad R \longrightarrow \infty, \label{bc_farfield_dimensionless} \\
\text{Initial condition:} \quad H = 0, \quad \textbf{A} = \textbf{I}  \quad &\text{at} \quad T = 0.  \label{bc_initial_dimensionless}
\end{align}
\end{subequations}
To further simplify the momentum equation (\ref{continuity_momentum_dimensionless}), it is useful to introduce a modified pressure variable above hydrostatic pressure and redefine the stress tensor to incorporate the conservative gravitational force,
\begin{equation}
\boldsymbol{\mathit{\Sigma}} = \boldsymbol{\Sigma} - \mathcal{G}Z\textbf{I}, \quad \mathcal{P} = P + \mathcal{G}Z.
\end{equation}
With these reformulated variables, the total stress equation in the fluid  (\ref{stress_dimensionless}) and the momentum equation (\ref{continuity_momentum_dimensionless}) simplify to
\begin{equation}\label{momentum_equation_dimensionless}
\boldsymbol{\mathit{\Sigma}} = -\mathcal{P} \textbf{I} + 2\beta_s \boldsymbol{\mathit{E}} + \frac{\beta_p}{Wi}(\textbf{A}-\textbf{I}), \quad 
\nabla \cdot \boldsymbol{\mathit{\Sigma}} = \textbf{0}.
\end{equation}
To account for the new variables, the stress boundary condition (\ref{normal_stress_bc}) becomes 
\begin{equation}\label{normal_stress_bc_dimensionless}
\textbf{n} \cdot \boldsymbol{\mathit{\Sigma}} = -\mathcal{G}Z \textbf{n}  \quad \text{at} \quad Z = H(R,T).
\end{equation}

\section{Preliminary setups}\label{setup}
\subsection{Distinguished limit}\label{distinguish}

We focus on two key non-dimensional parameters: the Weissenberg number, $Wi$, and the gravity parameter $\mathcal{G}$; we assume the viscosity ratio $\beta_p$ is given. It is natural to examine the asymptotic limit where $Wi \ll 1$ in which viscoelasticity appears as a correction to the Newtonian Stokes flow. However, careful consideration of the range of $\mathcal{G}$ is required, as different values of $\mathcal{G}$ will yield different asymptotic regimes depending on $Wi$.  The challenge of performing a double asymptotic expansion requires an understanding of the interaction between these parameters. Before advancing our analysis, we will investigate how key quantities, e.g., velocity, pressure, conformation tensor, and interface height, depend on $Wi$ and $\mathcal{G}$ to gain insight into the distinguished limit that forms the foundation of our study.

Initially, we assume the polymers are in equilibrium, with $\textbf{A} = \textbf{I}$. As the rod begins rotating at a constant angular velocity, the fluid forms a vortex around the rod, causing the suspended polymers to stretch.  Once the polymers extend beyond their equilibrium state, they induce additional polymeric stress of order $O(Wi)$, 
\begin{equation}\label{distinguish_polymerstress}
\boldsymbol{\mathit{\Sigma}}_{\text{polymer}} \thicksim Wi.
\end{equation}
When viscoelasticity is absent from the flow, the Stokes flow only yields a flat interface, i.e., in the absence of inertial effects. Therefore, we expect the normal stress near the interface to arise from the polymeric stress. Using the normal stress boundary condition (\ref{normal_stress_bc_dimensionless}), we find that the normal stress near the interface (deformation $H$) behaves as
\begin{equation}\label{distinguish_interfacestress}
\boldsymbol{\mathit{\Sigma}} \thicksim \boldsymbol{\mathit{\Sigma}}_{\text{polymer}} \thicksim \mathcal{G}H \quad (\text{near the interface}).
\end{equation}
By combining (\ref{distinguish_polymerstress}) and (\ref{distinguish_interfacestress}), we anticipate that the typical interface height will depend on both viscoelasticity and gravity as follows,
\begin{equation}\label{typical_interface_scaling}
    H \thicksim \frac{Wi}{\mathcal{G}}.
\end{equation}
As discussed earlier, the dynamics of the interface are driven by the (weak) viscoelasticity of the fluid. Therefore, we expect the characteristic time scale for temporal changes in the interface height to be determined by the polymer relaxation time, rather than the flow time scale itself. With this in mind, we use the kinematic boundary condition (\ref{kinematic_bc}) and the typical interface height  (\ref{typical_interface_scaling}) to estimate the typical magnitude of the upward fluid velocity near the interface as
\begin{equation}\label{distinguish_interface_velocity}
U_{\text{interface}} \thicksim \frac{H}{Wi} \thicksim \frac{1}{\mathcal{G}}.
\end{equation}      
Near the interface, we observe from (\ref{distinguish_polymerstress}) that the polymeric stress scales as $Wi$, whereas from (\ref{distinguish_interface_velocity}), the viscous stress scales like $\mathcal{G}^{-1}$. When $Wi \gg \mathcal{G}^{-1}$, the normal stress at the interface comes mainly from the polymeric stress as we had assumed to obtain (\ref{distinguish_interfacestress}). 

However, when $\mathcal{G} \approx Wi^{-1}$, the upward flow near the interface induces a viscous stress comparable to the polymeric stress, adding further complexity to the problem. The distinguished limit $\mathcal{G} \approx Wi^{-1}$, where our qualitative analysis ((\ref{distinguish_polymerstress})--(\ref{distinguish_interface_velocity})) begins to break down, motivates us to explore the gravity parameter $\mathcal{G}$ within the range $\mathcal{G} \gtrsim Wi^{-1}$. To simplify the complexity of the double asymptotic expansion into a single expansion in terms of $Wi$, based on the distinguished limit, we introduce a rescaled gravity parameter $\mathrm{G}$ by defining
\begin{equation}
\mathrm{G} = Wi\mathcal{G},
\end{equation}
where $\mathrm{G}$ is now taken to be of order $O(1)$. We anticipate that the problem remains solvable, at least when $\mathrm{G} \gg 1$. Intuitively, when gravitational effects are large, the viscoelastic fluid has difficulty climbing the rotating rod, which leads to only small deformations on the interface profile. This is precisely the small deformation situation that we aim to address in this study.

\subsection{Perturbation expansion in Weissenberg number}\label{Wi_perturbation_section}

In the following sections, we expand the velocity, pressure, and conformation tensor fields as regular perturbations in $Wi \ll 1$, 
\begin{subequations}\label{Wi_expansion}
\begin{align}
\mathcal{P} &= \mathcal{P}^{(0)} + Wi \mathcal{P}^{(1)} + ..., \\
\boldsymbol{\mathit{E}} &= \boldsymbol{\mathit{E}}^{(0)} + Wi \boldsymbol{\mathit{E}}^{(1)} + ..., \\
\textbf{A} &= \textbf{I}+ Wi \textbf{A}^{(1)} + Wi^2 \textbf{A}^{(2)} + ... \label{conformation_expansion} ,
\end{align}
\end{subequations}
where the superscript indicates the corresponding order in $Wi$.
A reader might observe that prescribing $\textbf{A} = \textbf{I}$ at the initial time (\ref{bc_initial_dimensionless}) and in the far field (\ref{bc_farfield_dimensionless}) does not necessarily guarantee that $\textbf{A} = \textbf{I}$ holds at leading order for all time. To ensure this, one must verify it through the evolution of the conformation tensor (\ref{conformation_dimensionless}). Nevertheless, we initially express the expansion this way to simplify the derivation of the interface boundary condition in the next subsection $\mathsection$~\ref{domain_perturbation}. This point will be revisited in $\mathsection$$\mathsection$~\ref{steady_section} and \ref{transient_section}.

Along with the quantities  expressed as expansions in (\ref{Wi_expansion}), we introduce an expansion of the interface profile in the situation where the interface is slightly deformed,
\begin{equation}\label{Wi_expansion_interface}
H(R,T) = Wi^2 H^{(2)}  + Wi^3 H^{(3)} + ...
\end{equation}
Although it is more natural to start the expansion of $H$ at order $O(Wi)$, since the interface remains flat in the absence of viscoelasticity, we have opted to express it this way for two reasons. First, we observe from our preliminary scaling argument (\ref{typical_interface_scaling}) that $H \thicksim Wi\mathcal{G}^{-1} \thicksim Wi^2 \mathrm{G}^{-1} = O(Wi^2)$. Second, as we will see in the following subsection $\mathsection$~\ref{domain_perturbation}, expanding the free-surface boundary condition to be consistent with the perturbation expansion in the Weissenberg number involves extensive calculations unless each term is scaled correctly. By keeping the non-zero leading-order expansion of $H$, the derivations for the boundary conditions at each order become more organized. In this work, we focus on obtaining an analytical formula for the leading-order term of the interface profile $H^{(2)}$, given in (\ref{Wi_expansion_interface}).

\subsection{Domain perturbation method for the interface boundary condition}\label{domain_perturbation} 

In $\mathsection$~\ref{scaling_section}, we introduced the normal-stress boundary condition (\ref{normal_stress_bc_dimensionless}) and the kinematic boundary condition (\ref{kinematic_bc}), in which both boundary conditions are evaluated at the interface $Z = H$. However, when expanding the interface profile $H$ in terms of $Wi$ (see \ref{Wi_expansion_interface}), it is challenging to evaluate the boundary conditions using the full series expansion. 
To resolve this issue, we take advantage of the assumption that the interface deformation is small to simplify the boundary conditions using a domain perturbation expansion. This step enables us to systematically write down the corresponding interface boundary conditions for the velocity, pressure, and conformation tensor fields at each order in $Wi$.

We start by expressing the normal stress boundary condition (\ref{normal_stress_bc_dimensionless}) for each order in $Wi$. The assumption of small deformation of the interface provides two simplifications. First, we can approximate the normal vector $\textbf{n}$ at the interface as $\textbf{n} \approx \textbf{e}_z - \frac{\partial H}{\partial R}\textbf{e}_r = \textbf{e}_z + O(Wi^2)$. Second, when evaluating the boundary condition at $Z = H$, the small value of $H$ allows us to perform a Taylor expansion around $Z = 0$. From these two remarks, we may simplify the left-hand side of (\ref{normal_stress_bc_dimensionless}) as,
\begin{align}\label{domain_perturbation_stress_lhs}
    \textbf{n} \cdot \boldsymbol{\mathit{\Sigma}}\bigg{|}_{Z = H} &= \textbf{n} \cdot \boldsymbol{\mathit{\Sigma}}\bigg{|}_{Z = 0} + H \frac{\partial (\textbf{n} \cdot \boldsymbol{\mathit{\Sigma}})}{\partial Z} \bigg{|}_{Z = 0} + \frac{1}{2}H^2 \frac{\partial^2 (\textbf{n} \cdot \boldsymbol{\mathit{\Sigma}})}{\partial Z^2} \bigg{|}_{Z = 0} + ... \\ 
    &=  \textbf{e}_z \cdot \left(-\mathcal{P}^{(0)}\textbf{I} + 2\beta_s \boldsymbol{\mathit{E}}^{(0)} + \beta_p \textbf{A}^{(1)}\right)\bigg{|}_{Z = 0} \notag \\ \nonumber
    &\: + Wi \left(\textbf{e}_z \cdot \left(-\mathcal{P}^{(1)}\textbf{I} + 2\beta_s \boldsymbol{\mathit{E}}^{(1)} + \beta_p \textbf{A}^{(2)}\right)\bigg{|}_{Z = 0}\right) + O(Wi^2).
\end{align}
On the other hand, the right-hand side of (\ref{normal_stress_bc_dimensionless}) is simply
\begin{equation}\label{domain_perturbation_stress_rhs}
-\mathcal{G}Z\textbf{n}\bigg{|}_{Z = H} = - \mathcal{G}H\textbf{e}_z + O(Wi^2) = -Wi \mathrm{G}H^{(2)}\textbf{e}_z + O(Wi^2).
\end{equation}
Combining (\ref{domain_perturbation_stress_lhs}) and (\ref{domain_perturbation_stress_rhs}), we conclude that  the stress boundary condition (\ref{normal_stress_bc_dimensionless}) at $O(1)$ and $O(Wi)$  becomes
\begin{subequations}
\begin{align}
\textbf{e}_z \cdot \left(-\mathcal{P}^{(0)}\textbf{I} + 2\beta_s \boldsymbol{\mathit{E}}^{(0)} + \beta_p \textbf{A}^{(1)}\right)\bigg{|}_{Z = 0}  &= \textbf{0}, \label{leading_order_stress_bc}\\
\textbf{e}_z \cdot \left(-\mathcal{P}^{(1)}\textbf{I} + 2\beta_s \boldsymbol{\mathit{E}}^{(1)} + \beta_p \textbf{A}^{(2)}\right)\bigg{|}_{Z = 0} &= -\mathrm{G}H^{(2)}\textbf{e}_z. \label{first_order_stress_bc}
\end{align}
\end{subequations}
At this stage, we choose not to express the kinematic boundary condition (\ref{kinematic_bc}) for each order in $Wi$. As mentioned in $\mathsection$~\ref{distinguish}, the temporal evolution of the interface height is expected to be governed by the polymer relaxation time, rather than the flow time scale. This implies that, starting from a flat interface, the interface will rise rapidly on a time scale of $O(Wi)$ before approaching its steady state. This behavior suggests a boundary layer-like structure in time, prompting the examination of two time scales: $T = O(1)$ and $T = O(Wi)$. In $\mathsection$~\ref{steady_section}, we focus on $T = O(1)$, representing the outer layer approximation, to recover the steady-state interface profile. In $\mathsection$~\ref{transient_section}, we analyze the shorter time scale $T = O(Wi)$ using a stretched time variable, corresponding to the inner layer approximation, to study the transient interface profile. We shall revisit the kinematic boundary condition (\ref{kinematic_bc}) separately for each rescaled time in $\mathsection$$\mathsection$~\ref{steady_section} and \ref{transient_section}.

\section{Outer layer: the steady-state interface 
profile}\label{steady_section}
\subsection{Leading-order solution}\label{steady_leading_section}
In this section, we examine the interface dynamics for $T = O(1)$, representing the state when time has progressed beyond the initial start-up of the flow. To this point, we have only provided a general qualitative outline explaining why a distinction in time scales is necessary. 
Now we examine the evolution of the conformation tensor by writing (\ref{conformation_dimensionless}) as
\begin{equation}\label{conformation_dimensionless_alt}
Wi\left(\frac{\partial \textbf{A}}{\partial T} +\textbf{U} \cdot \nabla \textbf{A} - \textbf{A} \cdot \nabla \textbf{U} - (\nabla \textbf{U})^{T}\cdot \textbf{A}\right) = -\left(\textbf{A}-\textbf{I}\right).
\end{equation}
Equation (\ref{conformation_dimensionless_alt}) features a first-order time derivative, with the highest-order term $\frac{\partial \textbf{A}}{\partial T}$ being multiplied by a small parameter $Wi$. This structure frequently arises in boundary-layer problems, where singular perturbation analysis is required to understand the solution behavior. Typically,  the analysis is split into two regions, one where (\ref{conformation_dimensionless_alt}) is used as is, and another where time is rescaled to maintain the dominance of the highest-order term; in this section we focus on the former case. Before proceeding further, we note that the right-hand side of (\ref{conformation_dimensionless_alt}) implies that, to leading order, $\textbf{A} = \textbf{I}$, regardless of the initial conditions. This is consistent with the expansion proposed in (\ref{conformation_expansion}). By substituting this expansion further, we express the first-order correction of the conformation tensor in terms of the leading-order velocity field,
\begin{equation}\label{conformation_first_order_steady}
    \textbf{A}^{(1)} = \nabla \textbf{U}^{(0)} + (\nabla \textbf{U}^{(0)})^{\rm T} = 2\boldsymbol{\mathit{E}}^{(0)}.
\end{equation}
In addition, when $T = O(1)$ and $H = O(Wi^2)$, we can derive the kinematic boundary conditions (\ref{kinematic_bc}) up to first-order terms, by employing a domain perturbation technique similar to that used in $\mathsection$~\ref{domain_perturbation}, as 
\begin{equation}\label{kinematic_steady}
    U_Z^{(0)}\big{|}_{Z = 0} = U_Z^{(1)}\big{|}_{Z = 0} = 0.
\end{equation}
To proceed with the problem solution, we substitute the first-order correction of the conformation tensor (\ref{conformation_first_order_steady}) into the leading-order momentum equation (\ref{momentum_equation_dimensionless}) to obtain 
\begin{equation}\label{Stokes_leading_order}
\textbf{0} = \nabla\cdot\left(-\mathcal{P}^{(0)} \textbf{I} + 2\beta_s \boldsymbol{\mathit{E}}^{(0)} + \beta_p\textbf{A}^{(1)}\right) = -\nabla\mathcal{P}^{(0)} + \nabla^2 \textbf{U}^{(0)}, 
\end{equation}where $\beta_s+\beta_p=1$ and from (\ref{conformation_first_order_steady}), we have 
$\nabla\cdot \textbf{A}^{(1)}= \nabla^2 \textbf{U}^{(0)}$.
Further, substituting (\ref{conformation_first_order_steady}) into the leading-order stress boundary condition (\ref{leading_order_stress_bc}) yields
\begin{equation}\label{leading_order_stress_bc_alt}
\textbf{e}_z \cdot \left(-\mathcal{P}^{(0)}\textbf{I} + 2 \boldsymbol{\mathit{E}}^{(0)} \right)\bigg{|}_{Z = 0}  = \textbf{0}.
\end{equation}
The leading-order governing equation (\ref{Stokes_leading_order}) is solved subject to the boundary conditions (\ref{bc_noslip_dimensionless}), (\ref{bc_farfield_dimensionless}), (\ref{kinematic_steady}), and (\ref{leading_order_stress_bc_alt}), which no longer depend on the viscoelastic effects. At this order, the problem simplifies to the rotation of a rod in a Newtonian fluid at low Reynolds number, which is known to produce an azimuthal vortex flow profile described by
\begin{equation}\label{vortex}
    \textbf{U}^{(0)} = \frac{1}{R}\textbf{e}_{\theta}, \quad\mathcal{P}^{(0)} = 0.
\end{equation}
With this leading-order velocity profile, we can write down the first-order correction to the conformation tensor,
\begin{equation}\label{first_order_conformation_steady}
\textbf{A}^{(1)} = 2\boldsymbol{\mathit{E}}^{(0)}= -\frac{2}{R^2}(\textbf{e}_{r}\textbf{e}_{\theta} + \textbf{e}_{\theta}\textbf{e}_{r}).
\end{equation}
Notice that  (\ref{first_order_conformation_steady}) does not satisfy the condition $\textbf{A}^{(1)} = \textbf{0}$ at the initial time; we will resolve this point in $\mathsection$~\ref{transient_section}. 

\subsection{First-order correction to the solution}\label{steady_first_section}

We now proceed to determine the first-order correction to the shape of the interface. From (\ref{first_order_conformation_steady}), we observe that the first-order correction to the conformation tensor appears as the rate-of-strain tensor from the leading-order vortical flow. This tensor has zero diagonal elements, resulting in no normal stress differences, which are typically responsible for the rod-climbing effect. To reveal the ``hoop stress" that leads to the rod-climbing effect, we solve for the second-order correction to the conformation tensor. Considering again the evolution of the conformation tensor equation (\ref{conformation_dimensionless_alt}), and using
(\ref{vortex}) and (\ref{first_order_conformation_steady}), we obtain
\begin{align}\label{second_order_conformation_steady}
    \textbf{A}^{(2)} &=  -\frac{\partial \textbf{A}^{(1)}}{\partial T}-\textbf{U}^{(0)}\cdot \nabla \textbf{A}^{(1)} + \textbf{A}^{(1)} \cdot \nabla\textbf{U}^{(0)}+(\nabla \textbf{U}^{(0)})^{T}\cdot \textbf{A}^{(1)} + 2\boldsymbol{\mathit{E}}^{(1)}  \notag \\
    &= \textbf{0} -\frac{4}{R^4}(\textbf{e}_{r}\textbf{e}_{r} - \textbf{e}_{\theta}\textbf{e}_{\theta}) + \frac{4}{R^4}(\textbf{e}_{r}\textbf{e}_{r} + \textbf{e}_{\theta}\textbf{e}_{\theta}) + 2\boldsymbol{\mathit{E}}^{(1)} \notag \\
    &= \frac{8}{R^4}\textbf{e}_{\theta}\textbf{e}_{\theta} + 2\boldsymbol{\mathit{E}}^{(1)}.
 \end{align}
 
 As in $\mathsection$~\ref{steady_leading_section}, we substitute (\ref{second_order_conformation_steady}) into the momentum equation (\ref{momentum_equation_dimensionless}) at first order, to find
\begin{align}\label{first_order_stokes}
\textbf{0} &= -\nabla \mathcal{P}^{(1)} + \beta_s \nabla^2 \textbf{U}^{(1)} + \beta_p \nabla \cdot \textbf{A}^{(2)} \notag \\
&= -\nabla \mathcal{P}^{(1)} + \nabla^2 \textbf{U}^{(1)} + \beta_p\nabla \cdot \left(\frac{8}{R^4}\textbf{e}_{\theta}\textbf{e}_{\theta}\right)\notag \\
  &= -\nabla \mathcal{P}^{(1)} + \nabla^2 \textbf{U}^{(1)} -\frac{8\beta_p}{R^5}\textbf{e}_r.
\end{align}
Observe from (\ref{second_order_conformation_steady}) that $\textbf{e}_z \cdot \textbf{A}^{(2)}=2\textbf{e}_z \cdot \boldsymbol{\mathit{E}}^{(1)}$. Thus, we can rewrite the stress boundary condition (\ref{first_order_stress_bc}) at first order as
\begin{equation}\label{first_order_stress_bc_alt}
\textbf{e}_z \cdot \left(-\mathcal{P}^{(1)}\textbf{I} + 2 \boldsymbol{\mathit{E}}^{(1)} \right)\bigg{|}_{Z = 0}  = -\mathrm{G}H^{(2)}\textbf{e}_z. 
\end{equation}

An important observation in solving the problem at first order is that all boundary conditions for the velocity field are constrained to be zero. Specifically, the no-slip boundary condition (\ref{bc_noslip_dimensionless}) imposes $\textbf{U}^{(1)} = \textbf{0}$ at $R = 1$, the far-field boundary condition (\ref{bc_farfield_dimensionless}) enforces $\textbf{U}^{(1)} = \textbf{0}$ as $R \longrightarrow \infty$, and the kinematic boundary condition (\ref{kinematic_steady}) gives $U_Z^{(1)}\big{|}_{Z = 0} = 0$. These features suggest  $\textbf{U}^{(1)} = \textbf{0}$, reducing the problem to solving for the pressure and interface height using (\ref{first_order_stokes}) and (\ref{first_order_stress_bc_alt}) with the corresponding boundary conditions. Specifically, (\ref{first_order_stokes}) implies that $\mathcal{P}^{(1)}(R)$, allowing us to simplify (\ref{first_order_stress_bc_alt}) to $\mathcal{P}^{(1)} = \mathrm{G}H^{(2)}$. The far-field boundary condition (\ref{bc_farfield_dimensionless}) then implies that the pressure decays to zero as the interface flattens, i.e., $\mathcal{P}^{(1)}(R \longrightarrow \infty) = \mathrm{G}H^{(2)}(R \longrightarrow \infty) = 0$. 
We can now integrate the momentum equation (\ref{first_order_stokes}) once with respect to $R$ to find the pressure, which then leads to the steady free-surface profile,
\begin{equation}
\mathcal{P}^{(1)} (R)= \frac{2\beta_p}{R^4}\qquad\hbox{and}\qquad \quad H^{(2)} (R)= \frac{2\beta_p}{\mathrm{G}R^4}.
\end{equation}
Thus, we can now approximate the free-surface profile at leading order as 
\begin{equation}\label{steady_interface_height}
    H_{\text{steady}}(R) \approx H^{(2)}Wi^2 = \frac{2\beta_p Wi^2}{\mathrm{G}R^4} = \frac{2\beta_p Wi}{\mathcal{G}R^4}.
\end{equation}

Before proceeding further, there are several points to highlight. First, we note that (\ref{steady_interface_height}) is consistent with our preliminary qualitative analysis in (\ref{typical_interface_scaling}), and that 
$\frac{d H_{\text{steady}}}{d R} = -\frac{8\beta_p Wi^2}{\mathrm{G}R^5} = O(Wi^2)$ is small. This confirms that our small-deformation assumption is valid. Second, the result  (\ref{steady_interface_height}) 
was obtained previously using an alternative second-order fluid model \citep{joseph1973,more2023rod}, which is known to agree with the Oldroyd-B model at steady state for small Weissenberg numbers. However, in earlier studies, domain perturbation methods were developed using dimensional rotation speeds, without explicitly addressing the impact of individual dimensionless parameters. In contrast, we present our analysis in dimensionless variables,  exposing the range of validity. In $\mathsection$$\mathsection$~\ref{transient_section} and \ref{with_inertia_section}, we increase the complexity by introducing time-dependent and inertia terms, while maintaining the same dimensionless analytical structure established here.

\section{Inner layer: the transient interface profile}\label{transient_section}
\subsection{Leading-order solution}\label{transient_leading_section}

In $\mathsection$~\ref{steady_section}, we have examined the problem at an intermediate time scale, $T = O(1)$, and observed that the leading-order interface profile (\ref{steady_interface_height}) is time-independent. This result occurs because, at $T = O(1)$, substituting the conformation tensor expansion (\ref{conformation_expansion}) into (\ref{conformation_dimensionless_alt}) has corrections (\ref{first_order_conformation_steady}) and (\ref{second_order_conformation_steady}) that do not satisfy the initial condition $\textbf{A} = \textbf{I}$ at $T = 0$. Physically, these features suggest that no transient changes in the conformation tensor are observed at intermediate times, as the dynamics of these changes occur on a much shorter time scale. Therefore, we introduce a stretched time variable to balance the time derivative term on the left-hand side of (\ref{conformation_dimensionless_alt}), $Wi \frac{\partial \textbf{A}}{\partial T}$, with the right-hand side of (\ref{conformation_dimensionless_alt}), $\textbf{A}-\textbf{I}$, by defining,
\begin{equation}\label{stretched_time}
    \tau = \frac{T}{Wi}.
\end{equation}In terms of physical variables, time has now been scaled by the relaxation time.
More precisely, we rescale time by interpreting $T = O(Wi)$ as $\tau = O(1)$. The evolution equation for the conformation tensor (\ref{conformation_dimensionless_alt}) in terms of the stretched time variable $\tau$ now becomes
\begin{equation}\label{conformation_stretched}
    Wi\left (\textbf{U} \cdot \nabla \mathsfbi{A} - \mathsfbi{A} \cdot \nabla \textbf{U} - (\nabla \textbf{U})^{\rm T}\cdot \mathsfbi{A}\right ) = -\left(\frac{\partial \mathsfbi{A}}{\partial \tau} + \mathsfbi{A}-\textbf{I}\right),
\end{equation}
where we distinguish the conformation tensor in the inner layer with $\mathsfbi{A}(R,Z,\tau)$. In (\ref{conformation_expansion}), we introduced an expansion for the conformation tensor, assuming the identity tensor $\textbf{I}$ as the leading-order term at all times. However, it is not immediately clear whether this assumption holds for any initial conditions. To verify this, we assume that the leading-order term of the conformation tensor is $\mathsfbi{A}^{(0)}$. Considering the leading-order terms of (\ref{conformation_stretched}), which is to be solved with  $\mathsfbi{A} = \textbf{I}$ at $\tau = 0$, 
we find
\begin{equation} 
\frac{\partial \mathsfbi{A}^{(0)}}{\partial \tau} = -(\mathsfbi{A}^{(0)} - \textbf{I}) \quad \Longrightarrow \quad \mathsfbi{A}^{(0)} = \textbf{I} \quad \forall \tau \geq 0. 
\end{equation}
If we had prescribed a pre-stretched conformation tensor away from equilibrium at the initial time, the leading-order term $\mathsfbi{A}^{(0)}$ would differ, though it would still relax back to the identity tensor as $\tau \longrightarrow \infty$. Nevertheless, assuming the identity tensor as the initial condition is reasonable for this problem, as no polymer stretching would occur before the rod begins rotating.

Following the approach in $\mathsection$~\ref{steady_leading_section}, we express the first-order correction to the conformation tensor in terms of the leading-order velocity profile using (\ref{conformation_stretched}) as
\begin{equation}\label{first_order_transient_conformation}
\frac{\partial \mathsfbi{A}^{(1)}}{\partial \tau} + \mathsfbi{A}^{(1)} = \nabla \textbf{U}^{(0)} + ( \nabla \textbf{U}^{(0)} )^{\rm T}= 2\boldsymbol{\mathit{E}}^{(0)} \quad \Longrightarrow \quad \mathsfbi{A}^{(1)} = 2e^{-\tau}\int_{0}^{\tau}\boldsymbol{\mathit{E}}^{(0)}e^{s}\:ds,
\end{equation}
where we have used the condition $\mathsfbi{A}^{(1)} = \textbf{0}$ at $\tau = 0$. We then substitute (\ref{first_order_transient_conformation}) into the leading-order momentum equation (\ref{momentum_equation_dimensionless}) to arrive at
\begin{equation}\label{transient_stokes_leading}
    \textbf{0} = -\nabla \mathcal{P}^{(0)} + \beta_s \nabla^2 \textbf{U}^{(0)} + \beta_p e^{-\tau}\int_0^{\tau} \nabla^2 \textbf{U}^{(0)}e^s ds.
\end{equation}
Equation (\ref{transient_stokes_leading}) is solved subject to the no-slip boundary condition (\ref{bc_noslip_dimensionless}), the far-field boundary condition (\ref{bc_farfield_dimensionless}), the stress boundary condition (\ref{leading_order_stress_bc}), and the kinematic boundary condition (\ref{kinematic_bc}). 

It is helpful to introduce a linear operator acting on the time-dependent vector field that resembles the right-hand-side of (\ref{transient_stokes_leading}) by defining 
\begin{equation}
\mathcal{L}(\textbf{F}(\textbf{X},\tau)) =  \beta_s \textbf{F}(\textbf{X},\tau) + \beta_p e^{-\tau} \int_{0}^{\tau} \textbf{F}(\textbf{X},s)e^{s}\:ds,
\end{equation}
so that (\ref{transient_stokes_leading}) becomes $\textbf{0} = -\nabla \mathcal{P}^{(0)} + \mathcal{L}\left (\nabla^2 \textbf{U}^{(0)}\right )$. With this notation, we simplify the stress boundary condition (\ref{leading_order_stress_bc}) by substituting (\ref{first_order_transient_conformation}) to obtain
\begin{equation}\label{leading_order_transient_kinematic}
2\mathcal{L}\left(\textbf{e}_z \cdot \boldsymbol{\mathit{E}}^{(0)}\bigg{|}_{Z = 0}\right) = \mathcal{P}^{(0)}\bigg{|}_{Z = 0} \textbf{e}_{z}.
\end{equation}

The kinematic boundary condition (\ref{kinematic_bc}), however, requires more careful examination, as it involves a time derivative that is crucial to the analysis in this section. Nevertheless, at leading order, we still find $U_Z^{(0)} \big|_{Z = 0} = 0$.
Since the flow begins at rest with a flat interface, this absence of upward velocity suggests that the interface will remain flat throughout.  Therefore, we aim to find a solution where $ U_Z^{(0)}$ is identically zero and the velocity field $ \textbf{U}^{(0)} = \textbf{U}^{(0)}(R, \tau) $ is independent of $Z$. The continuity equation (\ref{continuity_momentum_dimensionless}), expressed in cylindrical coordinates as $ \frac{1}{R} \frac{\partial (R U_R^{(0)})}{\partial R} + \frac{\partial U_Z^{(0)}}{\partial Z} = 0 $, then forces $ U_R^{(0)} = c(\tau)/{R} $ for some function $ c=c(\tau) $. Applying the no-slip boundary condition (\ref{bc_noslip_dimensionless}) implies $ c(\tau) \equiv 0 $, resulting in a purely azimuthal velocity field, $\textbf{U}^{(0)} = U_{\theta}^{(0)}(R, \tau) \textbf{e}_\theta$. 

Continuing this approach, the momentum equation (\ref{momentum_equation_dimensionless}), at leading order in the $r$- and $z$-directions, yields $\frac{\partial \mathcal{P}^{(0)}}{\partial R} = \frac{\partial \mathcal{P}^{(0)}}{\partial Z} = 0$, so that $ \mathcal{P}^{(0)}(\tau)$ only, while the stress boundary condition (\ref{leading_order_transient_kinematic}) in the $z$-direction further implies $\mathcal{P}^{(0)} \equiv 0 $. Thus, under the flat interface assumption, we have deduced that the pressure is absent from the leading-order solution. The momentum equation (\ref{momentum_equation_dimensionless}) and the stress boundary condition (\ref{leading_order_transient_kinematic}) simplify to 
\begin{equation}\label{leading-order-transient-reduced-equations}
    \mathcal{L}\left (\nabla^{2}\textbf{U}^{(0)}\right ) = \mathcal{L}\left(\textbf{e}_z \cdot \boldsymbol{\mathit{E}}^{(0)}\bigg{|}_{Z = 0}\right) = \textbf{0} \quad \Longrightarrow \quad \nabla^{2}\textbf{U}^{(0)} = \textbf{e}_z \cdot \boldsymbol{\mathit{E}}^{(0)}\bigg{|}_{Z = 0} = \textbf{0},
\end{equation}
where we have used the fact that the kernel of the linear operator $\mathcal{L}$ is null (see Appendix~\ref{kernel_of_L}). We observe that (\ref{leading-order-transient-reduced-equations}) has the same structure as the leading-order solution analyzed in  $\mathsection$~\ref{steady_leading_section}; in the weak-viscoelastic limit, the leading-order solution is effectively Newtonian. Although the momentum equation (\ref{transient_stokes_leading}) contains a time-dependent integral, the solution remains time-independent. Specifically, the velocity field retains the form of a vortex solution as in (\ref{vortex}),
\begin{equation}\label{vortex_transient}
\textbf{U}^{(0)} = \frac{1}{R}\textbf{e}_{\theta}.
\end{equation}
With this leading-order velocity profile, we can express the first-order correction to the conformation tensor, using (\ref{first_order_transient_conformation}), as 
\begin{equation}\label{first_order_transient_conformation_tensor}
\mathsfbi{A}^{(1)} = 2e^{-\tau}\int_{0}^{\tau}\boldsymbol{\mathit{E}}^{(0)}e^{s}\:ds = 2(1-e^{-\tau})\boldsymbol{\mathit{E}}^{(0)} = -\frac{2(1-e^{-\tau})}{R^2}(\textbf{e}_{r}\textbf{e}_{\theta} + \textbf{e}_{\theta}\textbf{e}_{r}).
\end{equation}
Starting from zero at $\tau=0$, $ \textbf{A}^{(1)} $ increases monotonically toward the steady state (\ref{first_order_conformation_steady}), approaching it at an exponential decay rate on the time scale of $ \tau $. This behavior confirms our remark at the start of $\mathsection$~\ref{transient_leading_section} that transient changes in the conformation tensor occur on a shorter time scale when $ T = O(Wi) $.

\subsection{First-order correction solution}\label{transient_first_section}

Hitherto, we have derived that, for an inner layer in time where $ T =O(Wi) $, the leading-order flow remains the same vortex solution as in the outer layer in time where $ T = O(1) $. Although the velocity fields are the same at the leading order, the first-order correction to the conformation tensor differs: in the inner layer, $\mathsfbi{A}^{(1)}$ varies with time, whereas it does not in the outer layer. Similarly, we anticipate the polymeric stress arising from $\mathsfbi{A}^{(2)}$, which defined the interface shape in $\mathsection$~\ref{steady_first_section}, to also be time-dependent, enabling us to capture the transient interface profile. By noting that the first-order correction to the conformation tensor obtained in (\ref{first_order_transient_conformation_tensor}) differs from (\ref{first_order_conformation_steady}) by a time-dependent factor of $(1-e^{-\tau})$, we can determine the time evolution of $\mathsfbi{A}^{(2)}$ using similar calculations as in (\ref{second_order_conformation_steady}),
\begin{equation}\label{second_order_conformation_transient_eq}
\frac{\partial \mathsfbi{A}^{(2)}}{\partial \tau} + \mathsfbi{A}^{(2)} = 2\boldsymbol{\mathit{E}}^{(1)} + \frac{8(1-e^{-\tau})}{R^4}\textbf{e}_\theta \textbf{e}_\theta.
\end{equation}
Multiplying both sides by the integrating factor $e^{\tau}$, we find $\mathsfbi{A}^{(2)}$ to be
\begin{equation}\label{second_order_conformation_transient}
\mathsfbi{A}^{(2)} = 2e^{-\tau}\int_{0}^{\tau}\boldsymbol{\mathit{E}}^{(1)}e^s\:ds + \frac{8(1-e^{-\tau}(\tau+1))}{R^4}\textbf{e}_\theta \textbf{e}_\theta,
\end{equation}
where we have used the initial condition $\mathsfbi{A}^{(2)} = \textbf{0}$ at $\tau = 0$ to write the lower limit of the integral. We then substitute (\ref{second_order_conformation_transient}) into the first-order momentum equation (\ref{momentum_equation_dimensionless}) to obtain,
\begin{align}\label{first_order_stokes_transient}
\textbf{0} &= -\nabla \mathcal{P}^{(1)} + \beta_s \nabla^2 \textbf{U}^{(1)} + \beta_p \nabla \cdot \mathsfbi{A}^{(2)}  \notag \\
&= -\nabla \mathcal{P}^{(1)} + \beta_s \nabla^2 \textbf{U}^{(1)} + \beta_p e^{-\tau}\int_{0}^{\tau} \nabla^2 \textbf{U}^{(1)}e^s \: ds -\frac{8\beta_p(1-(\tau+1)e^{-\tau})}{R^5}\textbf{e}_r \notag \\
&= -\nabla \mathcal{P}^{(1)} + \mathcal{L}\left (\nabla^2 \textbf{U}^{(1)}\right ) -\frac{8\beta_p(1-(\tau+1)e^{-\tau})}{R^5}\textbf{e}_r.
\end{align}
From (\ref{first_order_stokes_transient}), we observe that as the suspended polymers begin to stretch from their relaxed state, the elastic response generates a time-dependent force per volume given by $\frac{8\beta_p(1-(\tau+1)e^{-\tau})}{R^5}\textbf{e}_r$, which acts purely in the radial direction and is zero initially. This hoop stress is expected to drive the fluid radially inward toward the rod, which then, by conservation of mass, would lead the interface to rise, resulting in the rod-climb effect. Notably, the force induced by the hoop stress is conservative. Thus, it is convenient to define a new pressure variable that includes the polymeric stress by setting \(\mathrm{P}^{(1)} = \mathcal{P}^{(1)} - \frac{2\beta_p(1-(\tau+1)e^{-\tau})}{R^4}\). This substitution gives \(\nabla \mathrm{P}^{(1)} = \nabla \mathcal{P}^{(1)} + \frac{8\beta_p(1-(\tau+1)e^{-\tau})}{R^5} \textbf{e}_r\), which allows (\ref{first_order_stokes_transient}) to be rewritten as
\begin{equation}\label{first_order_stokes_transient_alt}
0 = -\nabla \mathrm{P}^{(1)} + \mathcal{L}\left (\nabla^2 \textbf{U}^{(1)}\right ).
\end{equation}  
In addition to the usual no-slip and far-field boundary conditions, we anticipate that the free-surface boundary conditions at the interface are particularly crucial for this first-order solution.  We substitute (\ref{second_order_conformation_transient}) into the first-order normal stress boundary condition (\ref{first_order_stress_bc}) and express the equations using the newly defined pressure variable, leading to the result,
\begin{equation}\label{stress_bc_first_transient_alt}
     -\mathrm{P}^{(1)}\bigg{|}_{Z = 0}\textbf{e}_z+ 2\mathcal{L}\left(\textbf{e}_z \cdot \boldsymbol{\mathit{E}}^{(1)}\bigg{|}_{Z = 0}\right) = \left(\frac{2\beta_p(1-(\tau+1)e^{-\tau})}{R^4}-\mathrm{G}H^{(2)}\right)\textbf{e}_z.
\end{equation}
On the other hand, recall from (\ref{Wi_expansion_interface}) that $H =O(Wi^2)$. This implies that when $T = O(Wi)$, we have $\frac{\partial H}{\partial T} = O(Wi)$. In other words, temporal variations in the interface profile will emerge in this first-order solution. Unlike for the steady interface profile (\ref{kinematic_steady}), the corresponding kinematic boundary condition (\ref{kinematic_bc}) at this order becomes
\begin{equation}\label{kinematic_first_transient}
    U_Z^{(1)}\bigg{|}_{Z=0} = \frac{\partial H^{(2)}}{\partial \tau}.
\end{equation}
It is difficult to obtain exact analytical expressions for the velocity field that satisfies (\ref{first_order_stokes_transient_alt}), (\ref{stress_bc_first_transient_alt}), and (\ref{kinematic_first_transient}) for arbitrary values of $\mathrm{G} = \mathcal{O}(1)$. However, when $\mathrm{G} \gg 1$, the method of dominant balance enables progress in determining the interface profile, with the dominant terms in this limit being the two on the right-hand side of (\ref{stress_bc_first_transient_alt}),
\begin{equation}\label{dominant_balance}
\mathrm{G} H^{(2)} \thicksim \frac{2\beta_p(1-(\tau+1)e^{-\tau})}{R^4} \quad\Longrightarrow \quad H^{(2)} \thicksim \frac{2\beta_p(1-(\tau+1)e^{-\tau})}{\mathrm{G} R^4} = O\left(\frac{1}{\mathrm{G}}\right).
\end{equation}

To confirm the consistency of this dominant balance, we need to assess the relative magnitudes of the pressure and velocities in terms of the parameter $\mathrm{G}$. From (\ref{dominant_balance}), the kinematic boundary condition (\ref{kinematic_first_transient}) implies that $U_Z^{(1)} = O\left(\frac{1}{\mathrm{G}}\right)$. According to the continuity equation (\ref{continuity_momentum_dimensionless}), $U_R^{(1)}$ must match the order of $U_Z^{(1)}$, giving $\textbf{U}^{(1)} = O\left(\frac{1}{\mathrm{G}}\right)$. This, in turn, implies from the momentum equation (\ref{first_order_stokes_transient_alt}) that $\mathrm{P}^{(1)}$ is also of order $O\left(\frac{1}{\mathrm{G}}\right)$. Consequently, both the pressure and the rate-of-strain on the left-hand side of (\ref{stress_bc_first_transient_alt}) are of order $O\left(\frac{1}{\mathrm{G}}\right)$, while the right-hand side remains of order $O(1)$. Therefore, as $\mathrm{G} \gg 1$, so that $1/\mathrm{G}$ is small, we conclude that the right-hand side forms a consistent dominant balance as desired. Thus, in the limit when $Wi \ll 1$ and $\mathrm{G} \gg 1$, we obtain the interface profile, 
\begin{equation}\label{transient_interface_shape}
H_{\text{transient}}(R,\tau) \approx H^{(2)}Wi = \frac{2\beta_p(1-(\tau+1)e^{-\tau})Wi^2}{\mathrm{G} R^4} = \frac{2\beta_p(1-(\tau+1)e^{-\tau})Wi}{\mathcal{G} R^4}.
\end{equation}
In $\mathsection$~\ref{steady_section}, we have derived a steady interface profile (\ref{steady_interface_height}) for $T = O(1)$, whereas in this section, we obtain a transient interface profile (\ref{transient_interface_shape}) for $T =O(Wi)$. To ensure consistency between the two solutions, we compare their limits in the overlapping region in time, 
\begin{equation}
    \lim_{\tau \longrightarrow \infty} H_{\text{transient}}(R,\tau) = \frac{2\beta_p Wi}{\mathcal{G} R^4} = H_{\text{steady}}(R,T).
\end{equation}
We thus confirm that the interface dynamics occur on a short timescale, with $T = O(Wi)$, and that the transient interface profile we derived approaches the steady interface profile at long times. 

\begin{figure}
\begin{center}
\includegraphics[scale=0.33]{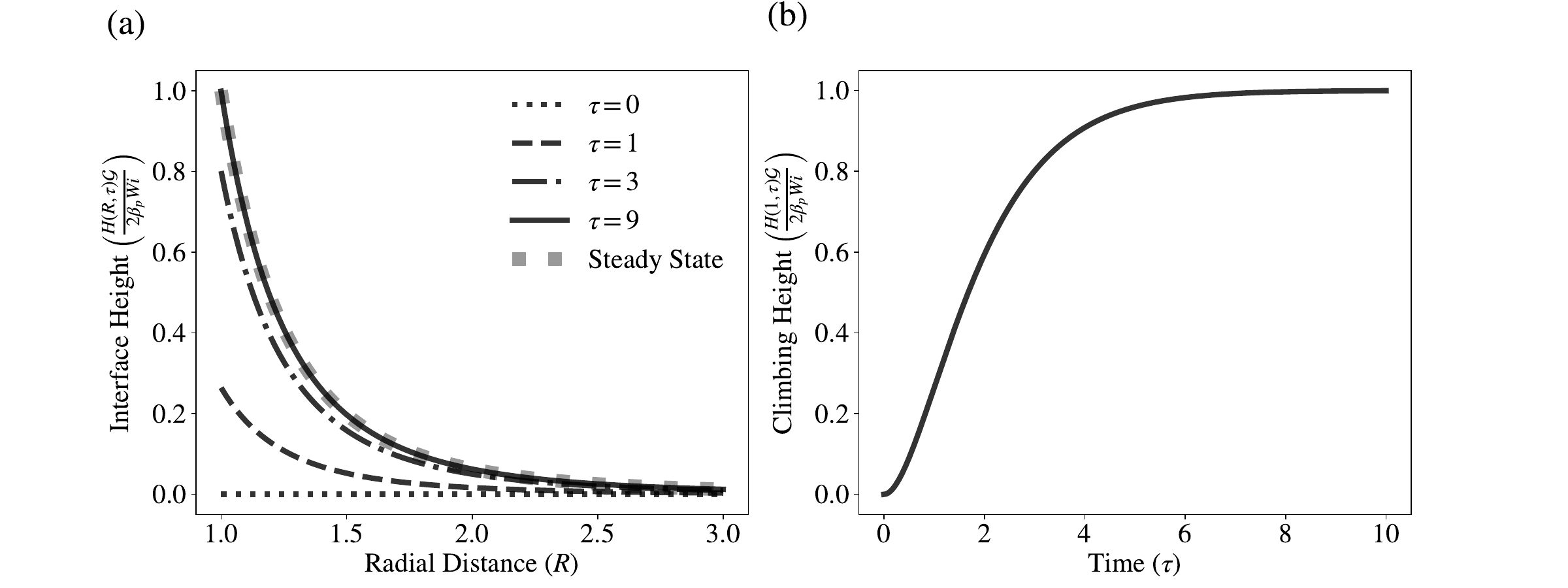}
\caption{($a$) Time evolution of the interface profile around an infinitely long rotating rod in an Oldroyd-B fluid, with inertial effects absent, shown at times $\tau = 0, 1, 3,$ and $10$. ($b$) Time evolution of the climbing height of an Oldroyd-B fluid on the rotating rod, with inertial effects absent, evaluated at $R = 1$.}
\label{interface_transient_plot}
\end{center}
\end{figure}

Figure \ref{interface_transient_plot}($a,b$) shows the time evolution of the scaled interface profile $H(R,\tau)\mathcal{G}/(2\beta_pWi)$ and the scaled the climbing height of an Oldroyd-B fluid on the rotating rod, $H(R=1,\tau)\mathcal{G}/(2\beta_pWi)$, given in~(\ref{transient_interface_shape}). We observe that the transient interface height reaches a steady state around $\tau \approx 6$, with the interface profile becoming nearly flat for $R \gtrapprox 3$.

\section{Inclusion of small but finite inertial effects}\label{with_inertia_section}
\subsection{Leading-order solution}

In $\mathsection$$\mathsection$~\ref{steady_section} and \ref{transient_section}, we excluded inertial effects to isolate the interplay between gravity and viscoelasticity. As derived in $\mathsection$~\ref{transient_section}, the transient interface profile (\ref{transient_interface_shape}) remains positive, owing to the hoop stress in the viscoelastic fluids, which drives the fluid radially inward and is responsible for rod-climbing. However, introducing a small inertial effect may change this dynamic, as the resulting centrifugal force would push the fluid radially outward. This competition between forces leads us to investigate how finite, small inertia may impact the interface height when both effects are present. In this section, we continue to use the stretched time scale $\tau$ to examine the transient interface dynamics.

\begin{figure}
\begin{center}
\includegraphics[scale=0.33]{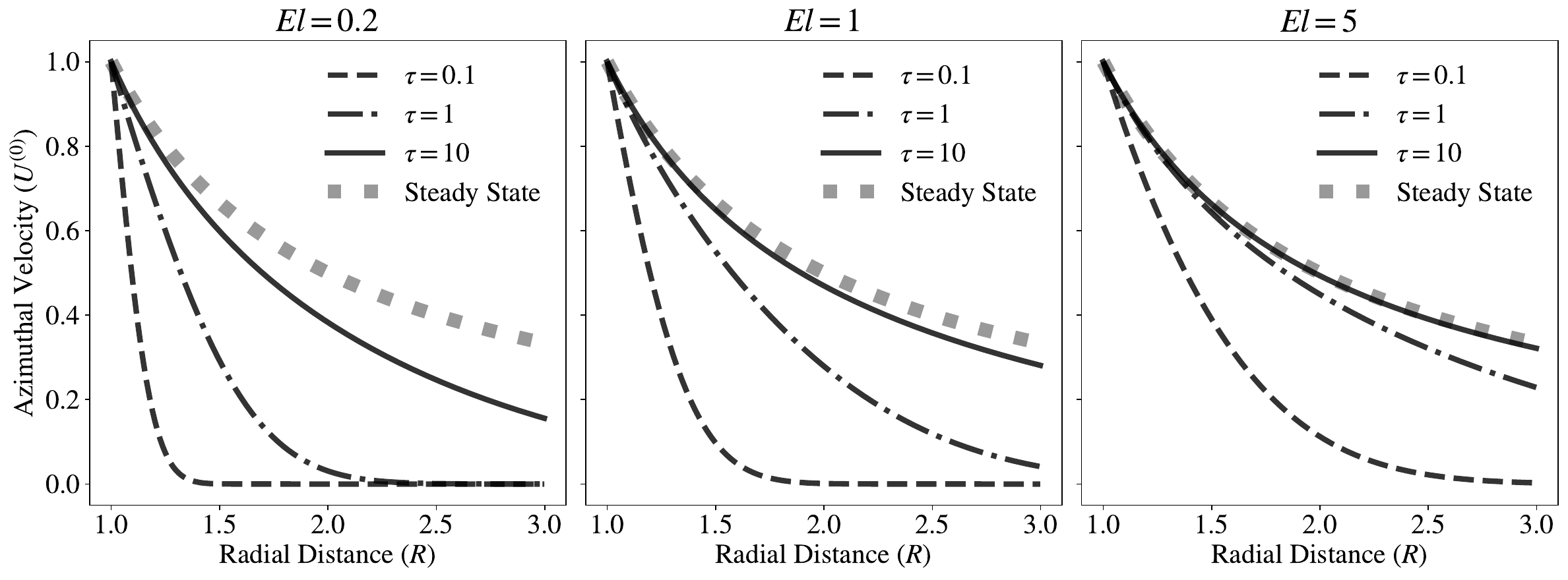}
\caption{Time evolution of the leading-order velocity field $U^{(0)}(R, \tau) $ around an infinitely long rotating rod in an Oldroyd-B fluid, incorporating small but finite inertial effects, shown at times $\tau = 0.1, 1,$ and $10$. We set $\beta_p = 0.5$ and explore three different values of the elasticity number $El= 0.2, 1,$ and $5$.}
\label{velocity_transient_plot}
\end{center}
\end{figure}

Incorporating inertial effects into our analysis, the momentum equation (\ref{momentum_equation_dimensionless}) is
\begin{equation}\label{momentum_equation_inertia}
Re \left(\frac{1}{Wi}\frac{\partial \textbf{U}}{\partial \tau} + \textbf{U}\cdot \nabla\textbf{U}\right) = \nabla \cdot \boldsymbol{\mathit{\Sigma}}.
\end{equation}
Rather than focusing our analysis on the Reynolds number, we find it more useful to base our analysis on the elasticity number, $El$, which represents the ratio of elastic stress to inertial stress \citep{DENN1971},
\begin{equation} 
El = \frac{Wi}{Re}. 
\end{equation}
Consequently, the momentum equation (\ref{momentum_equation_inertia}) can be expressed as
\begin{equation}\label{momentum_equation_inertia_alt}
El^{-1} \left(\frac{\partial \textbf{U}}{\partial \tau} + Wi \textbf{U}\cdot \nabla\textbf{U}\right) = \nabla \cdot \boldsymbol{\mathit{\Sigma}}.
\end{equation}
Previously, we derived the interface profile for $El \longrightarrow \infty$ and now aim to extend our analysis to cases where $El$ is large. To isolate the nonlinear term $\textbf{U} \cdot \nabla \textbf{U}$ from the leading-order velocity profile, we assume that $El$ is at least  $O(1)$, indicating small yet finite inertial effects with $Re \lessapprox Wi$. When $El$ is finite, the time-derivative $\frac{\partial \textbf{U}}{\partial \tau}$ of the velocity field reappears, in contrast to the low-Reynolds-number approximation in $\mathsection$$\mathsection$~\ref{steady_section} and \ref{transient_section}. To ensure the problem is well-posed, we specify the velocity field at the initial time,
\begin{equation}\label{velocity_initial_time}
    \textbf{U} = \textbf{0} \quad \text{at} \quad \tau = 0.
\end{equation}
Besides the additional terms on the left-hand side of the momentum equation (\ref{momentum_equation_inertia_alt}) and the new initial condition (\ref{velocity_initial_time}), the remaining constitutive equations and boundary conditions are identical to those in $\mathsection$~\ref{transient_section}. Specifically, the first-order correction to the conformation tensor, $\mathsfbi{A}^{(1)}$, can be expressed in terms of the leading-order rate-of-strain tensor, as shown in (\ref{first_order_transient_conformation}). This expression, combined with the momentum equation (\ref{momentum_equation_inertia_alt}), allows us to derive the differential equation for the leading-order velocity field,
\begin{equation}\label{inertia_stokes_leading}
    El^{-1}\frac{\partial \textbf{U}^{(0)}}{\partial \tau} = -\nabla \mathcal{P}^{(0)} + \mathcal{L}\left(\nabla^2 \textbf{U}^{(0)}\right).
\end{equation}
Following the approach in $\mathsection$~\ref{transient_leading_section}, we seek a solution where the leading-order interface is flat. Using the same reasoning, we deduce that $\mathcal{P}^{(0)} = 0$ and the velocity field is purely azimuthal, $ \textbf{U}^{(0)} = U^{(0)}(R,\tau) \textbf{e}_{\theta}$. This allows us to reduce the vector differential equation (\ref{inertia_stokes_leading}) to a partial differential equation with a single unknown variable,
\begin{equation}\label{inertia_stokes_leading_alt}
    El^{-1}\frac{\partial U^{(0)}}{\partial \tau} = \mathcal{L}\left(\frac{1}{R} \frac{\partial}{\partial R} \left(R\frac{\partial U^{(0)}}{\partial R}\right) - \frac{U^{(0)}}{R^2}\right).
\end{equation}
Equation (\ref{inertia_stokes_leading_alt}) is formulated on the semi-infinite domain $R \in [1, \infty)$ subject to the no-slip boundary condition (\ref{bc_noslip_dimensionless}), the far-field boundary condition (\ref{bc_farfield_dimensionless}), and the initial condition (\ref{velocity_initial_time}), i.e., 
\begin{equation}\label{1D_IC_BC}
    U^{(0)}(1,\tau) = 1, \quad U^{(0)}(\infty,\tau) = 0, \quad U^{(0)}(R,0) = 0.
\end{equation}
Since the vortex solution $\frac{1}{R}$ represents the steady-state solution  of (\ref{inertia_stokes_leading_alt}) along with the specified no-slip and far-field boundary conditions, we decompose $U^{(0)}$ into a steady-state component and a transient component, 
\begin{equation}
    U^{(0)}(R,\tau) = \frac{1}{R} + \tilde{U}^{(0)}(R,\tau).
\end{equation}
As (\ref{inertia_stokes_leading_alt}) is linear, $\tilde{U}^{(0)}$ also satisfies (\ref{inertia_stokes_leading_alt}) but with homogeneous spatial boundary conditions,
\begin{equation}\label{1D_IC_BC_alt}
    \tilde{U}^{(0)}(1,\tau) = 0, \quad \tilde{U}^{(0)}(\infty,\tau) = 0, \quad \tilde{U}^{(0)}(R,0) = -\frac{1}{R}.
\end{equation}
We approach this problem using a Weber transform $\mathcal{W}$ of order one \citep{watson1966bessel, piessens2000hankel}, an analog of the Hankel transform on the semi-infinite domain $[1, \infty)$, by introducing 
\begin{equation}\label{Weber_transform}
    V[\tau; k] = \mathcal{W}\left\{\tilde{U}^{(0)}(R, \tau)\right\} = \int_{1}^{\infty} R \tilde{U}^{(0)}(R, \tau) \, \mathit{Z}_{1}(k, R) \, dR,
\end{equation}
with the kernel $\mathit{Z}_{1}(k, R)$ given by  
\begin{equation}\label{Weber_transform_kernel}
    \mathit{Z}_{1}(k, R) = \mathit{J}_{1}(kR) \mathit{Y}_{1}(k) - \mathit{Y}_{1}(kR) \mathit{J}_{1}(k),
\end{equation}
where $\mathit{J}_1$ and $\mathit{Y}_1$ are the first- and second-kind Bessel functions of order one, respectively. This transformation enables us to simplify the Bessel differential operator of order one acting on $\tilde{U}^{(0)}$ on the right-hand side of (\ref{inertia_stokes_leading_alt}), provided that  $\tilde{U}^{(0)}$ decays to zero at infinity,
\begin{equation}
    \mathcal{W}\left\{\frac{1}{R} \frac{\partial}{\partial R} \left(R\frac{\partial \tilde{U}^{(0)}}{\partial R}\right) - \frac{\tilde{U}^{(0)}}{R^2}\right\} = -k^2 \mathcal{W}\left\{\tilde{U}^{(0)}(R, \tau)\right\} -  \cancelto{0}{\frac{2}{\pi}\tilde{U}^{(0)}(1,\tau)} = -k^2V[\tau;k].
\end{equation}
Note that the operator $\mathcal{W}$ acts on the $R$ variable, while $\mathcal{L}$ acts on $\tau$, which implies that the two operators commute: $\mathcal{W} \circ \mathcal{L} = \mathcal{L} \circ \mathcal{W}$. We apply the Weber transform to (\ref{inertia_stokes_leading_alt}) yielding,
\begin{equation}\label{inertia_stokes_leading_transformed}
    El^{-1}\frac{\partial V[\tau; k]}{\partial \tau} = -k^2\mathcal{L}\left(V[\tau;k]\right) = -k^2\left(\beta_s V[t;k] + \beta_p e^{-\tau}\int_{0}^{\tau} V[s;k] e^s \:ds\right).
\end{equation}
In this transformed variable, the initial condition becomes,
\begin{align}\label{inertia_inital_condition_transformed}
    V[0;k] &= \mathcal{W}\left\{\tilde{U}^{(0)}(R,0)\right\} = -\int_{1}^{\infty} \mathit{Z}_{1}(k,R) \: dR \notag \\
    &= \mathit{J}_{1}(k)\int_{1}^{\infty} \mathit{Y}_{1}(kR) \: dR -  \mathit{Y}_{1}(k)\int_{1}^{\infty}\mathit{J}_{1}(kR) \: dR \notag \\ 
    &= \frac{1}{k}\left(\mathit{J}_{1}(k)\mathit{Y}_{0}(k) -  \mathit{J}_0(k)\mathit{Y}_{1}(k)\right) = \frac{2}{\pi k^2},
\end{align}
where the last equality follows from the Wronskian of the zero-order Bessel differential equation. To solve (\ref{inertia_stokes_leading_transformed}), we first multiply both sides by $e^{\tau}$ and then differentiate with respect to $\tau$, to obtain
\begin{equation}\label{inertia_stokes_leading_transformed_alt}
    El^{-1}\left(\frac{\partial^2 V}{\partial \tau^2} + \frac{\partial V}{\partial \tau}\right) = -k^2\left(\beta_s \frac{\partial V}{\partial \tau} + V\right),
\end{equation}
where we have used the relation $\beta_s + \beta_p = 1$.
Equation (\ref{inertia_stokes_leading_transformed_alt}) is a constant-coefficient second-order differential equation in time, which has solutions of the form,
\begin{equation}\label{inertia_stokes_leading_transformed_sol_structure}
V[\tau;k] = C_1(k) e^{\lambda_{+} \tau} + C_2(k) e^{\lambda_{-} \tau},
\end{equation}
where $\lambda_{+}$ and $\lambda_{-}$ are roots of the equations
\begin{equation}\label{inertial_roots}
    \lambda^2 + \left(1 + k^2 El \beta_s\right)\lambda + k^2 El = 0.
\end{equation}
In particular,
\begin{equation}
    \lambda_{\pm} = \frac{-1-k^2 El \beta_s \pm \sqrt{(1+k^2 El \beta_s)^2 - 4k^2 El}}{2}.
\end{equation}
Although $\lambda_{\pm}$ can be complex depending on the value of $k$, their real parts are always negative, except at $k=0$, where one root is zero and the other is $-1$. In addition to the initial condition (\ref{inertia_inital_condition_transformed}), evaluating (\ref{inertia_stokes_leading_transformed}) at $\tau = 0$ provides an initial condition for $\frac{\partial V}{\partial \tau}$,
\begin{equation}\label{inertial_initial_condition_derivative_transformed}
    \frac{\partial V}{\partial \tau}[0;k] = -k^2 El \beta_sV[0;k] = -\frac{2 El\beta_s}{\pi }.
\end{equation}
Using (\ref{inertia_inital_condition_transformed}), (\ref{inertia_stokes_leading_transformed_sol_structure}), and (\ref{inertial_initial_condition_derivative_transformed}), and simplifying with (\ref{inertial_roots}), we find
\begin{equation}
    V[\tau;k] = \frac{2}{\pi}\left(\frac{\left(1+\lambda_{+}\right)e^{\lambda_{+} \tau} - \left(1 + \lambda_{-}\right) e^{\lambda_{-}\tau}}{k^2(\lambda_{+} - \lambda_{-})}\right),
\end{equation}
where $\lambda_{+}$ and $\lambda_{-}$ are functions of $k,El$, and $\beta_s$. Finally, we make an inverse Weber transform to obtain
\begin{align}\label{inertia_transient_sol}
    \mbox{\hspace{-5mm}}\tilde{U}^{(0)}(R, \tau) &= \int_{0}^{\infty} \frac{kV[\tau;k]\mathit{Z}_1(k,R)}{\mathit{J}_1^2(k) + \mathit{Y}_1^2(k)}\: dk \notag \\
    &= \frac{2}{\pi}\int_{0}^{\infty} \frac{\left(\left(1+\lambda_{+}\right)e^{\lambda_{+} \tau} - \left(1 + \lambda_{-}\right) e^{\lambda_{-} \tau}\right) \mathit{Z}_1(k,R)}{ k(\lambda_{+} - \lambda_{-})\left(\mathit{J}_1^2(k) + \mathit{Y}_1^2(k)\right)}\: dk. 
\end{align}
Although proceeding with (\ref{inertia_transient_sol}) analytically is difficult, it is straightforward to handle numerically. In figure~\ref{velocity_transient_plot}, we show the time evolution of the leading-order velocity profile $U^{(0)}(R,\tau)$ for various values of $El$. We observe that larger $El$ leads to a faster approach to a steady state in the transient velocity profile. Intuitively, when $El$ is large, the elastic effect ($Wi$) is larger than the inertial effect ($Re$). A larger viscoelasticity accelerates the flow's return to equilibrium by emphasizing the elastic ``spring-like" properties over inertial resistance, thus pushing the flow to a steady state faster.

\subsection{First-order correction solution}

In $\mathsection$$\mathsection$~\ref{steady_first_section} and \ref{transient_first_section}, we used the leading-order velocity field $\textbf{U}^{(0)}$ to calculate the polymeric force $\nabla \cdot \mathsfbi{A}^{(2)}$, which is then incorporated into the pressure gradient $\nabla \mathcal{P}^{(1)}$ to determine the transient interface shape. However, the leading-order velocity field (\ref{inertia_transient_sol}) is challenging to work with directly. To address this, we first express $\nabla \cdot \mathsfbi{A}^{(2)}$ in terms of $\textbf{U}^{(0)}$, calculate the interface profile, and then transition to numerical computations as a final step. We find (see Appendix~\ref{hoop_stress_calculation} for more detailed calculations)
\begin{align}\label{inertia_transient_polymeric_force}
    \nabla \cdot \mathsfbi{A}^{(2)} &= e^{-\tau}\int_{0}^{\tau} \nabla^2 \textbf{U}^{(1)}e^{s}\: ds \notag \\
    &\:- 2e^{-\tau}\int_{0}^{\tau}R\frac{\partial}{\partial R}\left(\frac{U^{(0)}(R,S)}{R}\right)\left(\int_{0}^{S}\frac{\partial}{\partial R}\left(\frac{U^{(0)}(R,s)}{R}\right)e^{s}\:ds\right)dS \:  \textbf{e}_{r}.
\end{align}
At the first order, the momentum equation (\ref{momentum_equation_inertia_alt}) becomes
\begin{equation}\label{inertia_transient_momentum_first}
El^{-1}\left(\frac{\partial \textbf{U}^{(1)}}{\partial \tau} + \textbf{U}^{(0)}\cdot \nabla\textbf{U}^{(0)}\right) = -\nabla \mathcal{P}^{(1)} + \beta_s \nabla^2 \textbf{U}^{(1)} + \beta_p \nabla \cdot \mathsfbi{A}^{(2)}.
\end{equation}
The second term on the left-hand side of (\ref{inertia_transient_momentum_first}) represents the centrifugal force generated by the (start-up) vortex,
\begin{equation}\label{centrifugal_force}
\textbf{U}^{(0)}\cdot \nabla\textbf{U}^{(0)} = -\frac{(U^{(0)})^2}{R}\textbf{e}_r.
\end{equation}
Observe that both the centrifugal force (\ref{centrifugal_force}) and the polymeric force induced by the leading-order flow (second term of (\ref{inertia_transient_polymeric_force})) are directed radially. To simplify, we introduce a modified pressure variable that combines these two conservative forces,
\begin{align}\label{inertia_transient_modified_pressure}
    \mathrm{P}^{(1)}(R,\tau) &= \mathcal{P}^{(1)}(R,\tau) + El^{-1} \int_{R}^{\infty} \frac{(U^{(0)}(\tilde{R},\tau))^2}{\tilde{R}} \: d\tilde{R} \notag \\ 
    &\:- 2\beta_p e^{-\tau}\int_{R}^{\infty}\int_{0}^{\tau}\tilde{R}\frac{\partial}{\partial \tilde{R}}\left(\frac{U^{(0)}(\tilde{R},S)}{\tilde{R}}\right)\left(\int_{0}^{S}\frac{\partial}{\partial \tilde{R}}\left(\frac{U^{(0)}(\tilde{R},s)}{\tilde{R}}\right)e^{s}\:ds\right)dSd\tilde{R}.
\end{align}
Combining (\ref{inertia_transient_momentum_first}) and (\ref{inertia_transient_modified_pressure}) allows us to express the momentum equation as
\begin{equation}\label{inertia_stokes_first}
    El^{-1}\frac{\partial \textbf{U}^{(1)}}{\partial \tau} = -\nabla \mathrm{P}^{(1)} + \mathcal{L}\left(\nabla^2 \textbf{U}^{(1)}\right).
\end{equation}
Following a similar argument to that in $\mathsection$~\ref{transient_first_section}, in the limit where $\mathrm{G} \gg 1$, we argue through the dominant balance that the interface profiles are determined by the pressure contributions from the inertial and polymeric stresses in (\ref{inertia_transient_modified_pressure}). We obtain
\begin{align}\label{inertia_interface_profile}
   H_{\text{inertia}}(&R, \tau) \approx -\frac{Re}{\mathcal{G}} \int_{R}^{\infty} \frac{(U^{(0)}(\tilde{R},\tau))^2}{\tilde{R}} \: d\tilde{R} \notag\\
   &\:+ \frac{2\beta_p Wi e^{-\tau}}{\mathcal{G}} \int_{R}^{\infty}\int_{0}^{\tau}\tilde{R}\frac{\partial}{\partial \tilde{R}}\left(\frac{U^{(0)}(\tilde{R},S)}{\tilde{R}}\right)\left(\int_{0}^{S}\frac{\partial}{\partial \tilde{R}}\left(\frac{U^{(0)}(\tilde{R},s)}{\tilde{R}}\right)e^{s}\:ds\right)dSd\tilde{R}. 
\end{align}
\begin{figure}
\begin{center}
\includegraphics[scale=0.33]{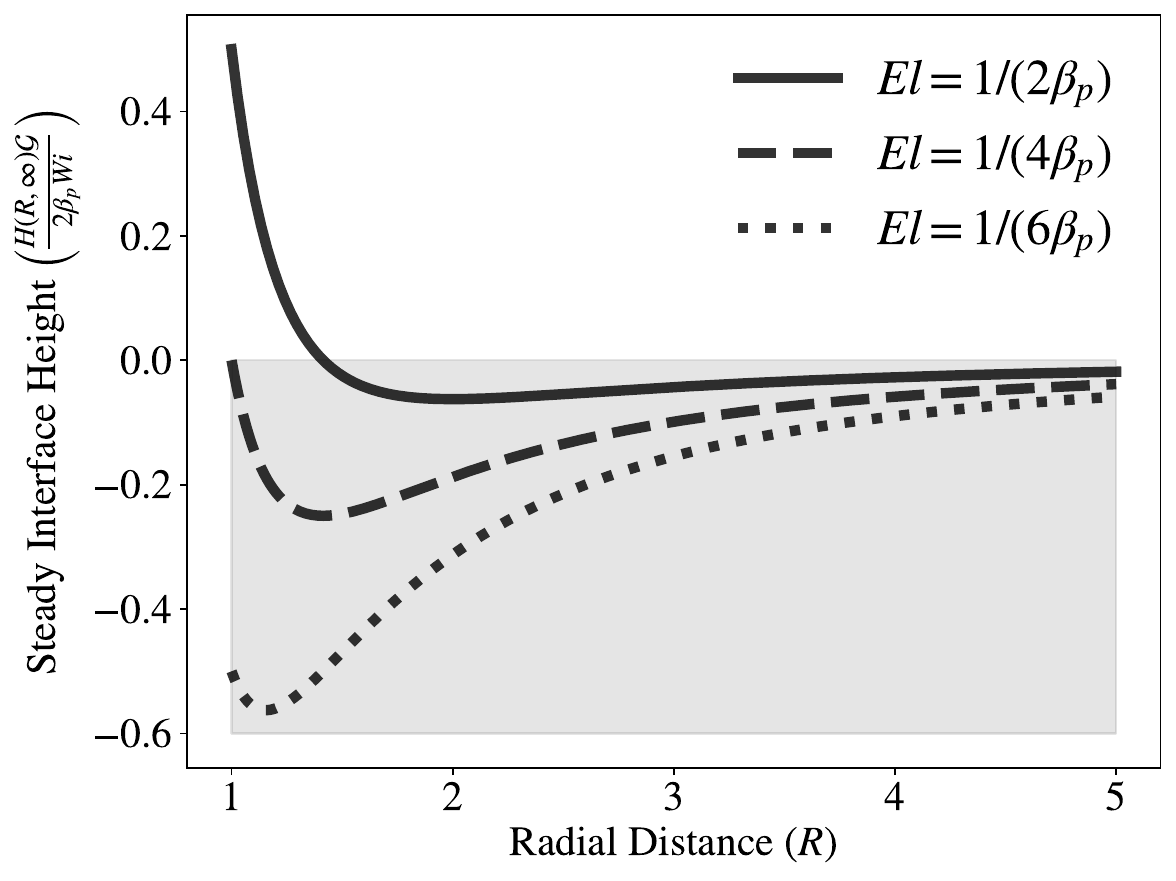}
\caption{Steady-state interface profiles around an infinitely long rotating rod in an Oldroyd-B fluid, incorporating small but finite inertial effects, shown for elasticity numbers $El = 1/(2\beta_p), 1/(4\beta_p)$, and $1/(6\beta_p)$. }
\label{inertia_steady_plot}
\end{center}
\end{figure}
Before we present numerical results to (\ref{inertia_interface_profile}), we consider the steady interface profile in the presence of inertial effects. We substitute $U^{(0)} \approx \frac{1}{R}$ and let $\tau \longrightarrow \infty$ in (\ref{inertia_interface_profile}) to find 
\begin{align}\label{interface_inertia_steady_state}
    H_{\text{inertia,\ steady}}(R) 
    \approx -\frac{Re}{2\mathcal{G}R^2}+ \frac{2\beta_p Wi}{\mathcal{G}R^4}. 
\end{align}
At the rod ($R = 1$), the steady rise height is given by $-\frac{Re}{2\mathcal{G}} + \frac{2\beta_p Wi}{\mathcal{G}}$. Rod-climbing, at the steady state, occurs when this rise height is positive, which corresponds to
\begin{equation}\label{climbing-criterion}
    Re < 4\beta_p Wi \quad \Longleftrightarrow \quad 4El \beta_p > 1.
\end{equation}
This result agrees with the predictions of the second-order fluid model \citep{joseph1973,bird1987dynamics1}, assuming that the first normal stress coefficient is $\Psi_1 = 2\mu_p \lambda$ and the second normal stress coefficient is $\Psi_2 = 0$. The second-order fluid model additionaly requires the climbing constant $\beta = \frac{1}{2}(\Psi_1 + 4\Psi_2)$ to satisfy $\beta > 0$, a condition that is automatically satisfied in the case of the Oldroyd-B model.

The climbing criterion (\ref{climbing-criterion}) requires viscoelastic effects (represented by $Wi$) to surpass inertial effects (represented by $Re$) for climbing to occur, though the effective viscoelasticity is also reduced by the polymer solution's diluteness (represented by $\beta_p$). This highlights how the balance between viscoelasticity, inertia, and polymer concentration governs the climbing behavior. Figure~\ref{inertia_steady_plot} illustrates the steady interface profile, $H_{\text{inertia,\ steady}}$, scaled by $2\beta_p Wi/\mathcal{G}$, for different values of $El$. A positive interface height near the rod is observed for $El > 1/(4\beta_p)$, whereas a negative interface height is observed for $El < 1/(4\beta_p)$. 

\begin{figure}
\begin{center}
\includegraphics[scale=0.33]{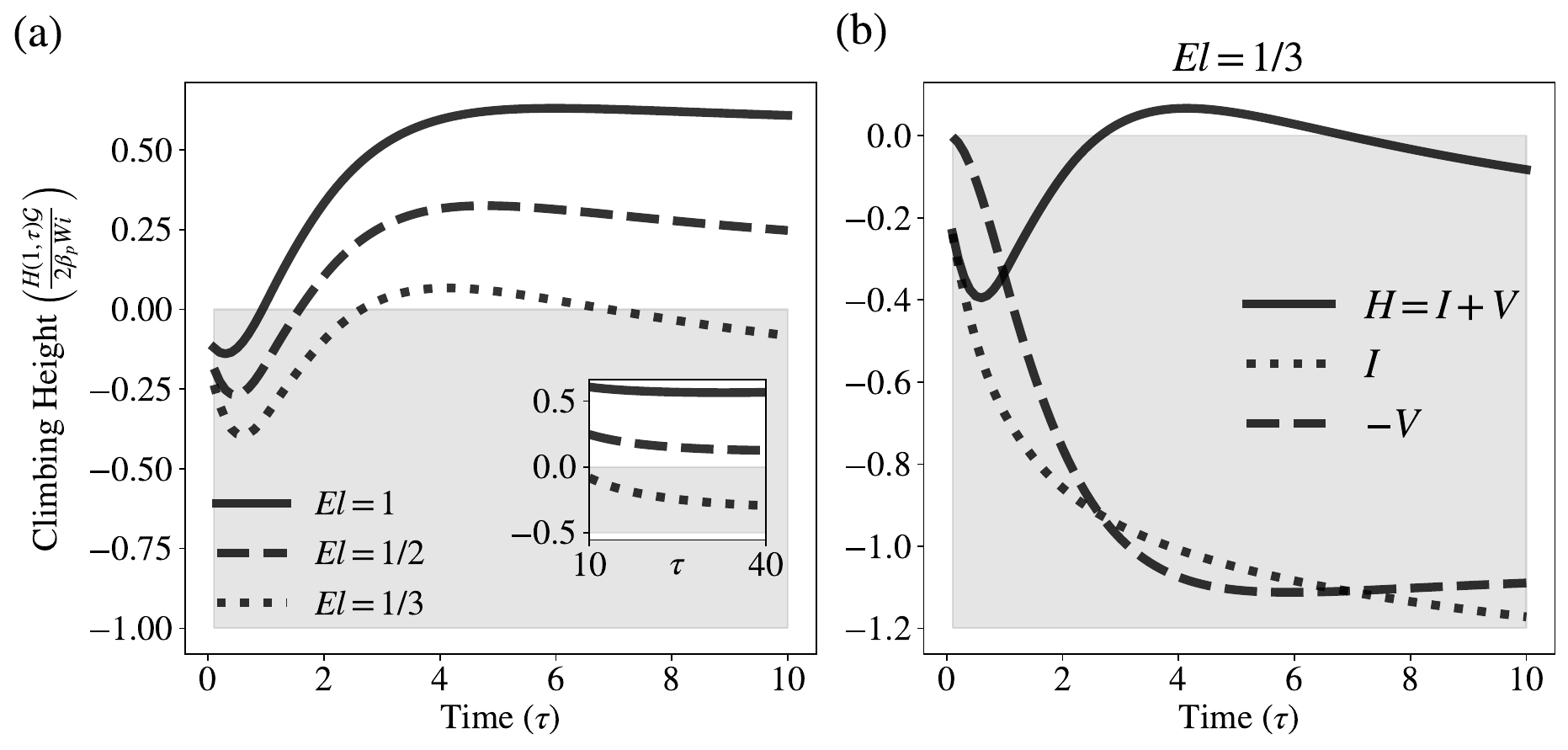}
\caption{($a$) Time evolution of the climbing height of an Oldroyd-B fluid at $R = 1$ for elasticity numbers $El = 1$, $1/2$, and $1/3$, with $\beta_p = 0.5$. ($b$) Breakdown of contributions to the climbing height of viscoelastic fluids at $R = 1$, for $El = 1/3$ and $\beta_p = 0.5$, where $I$ denotes the inertial contribution and $V$ denotes the viscoelastic contribution.}
\label{inertia_climbing_plot}
\end{center}
\end{figure}

To examine the interface dynamics before reaching the steady state (\ref{interface_inertia_steady_state}), we present the transient interface (\ref{inertia_interface_profile}) obtained via numerical computations. In figure~\hyperref[inertia_steady_plot]{\ref{inertia_climbing_plot}($a$)}, we examine the climbing height (scaled by $2\beta_p Wi/\mathcal{G}$) of a viscoelastic Oldroyd-B fluid on a rotating rod at $R = 1$ for various elasticity numbers. Initially, the impulsive motion of the rod causes a sudden dip in the fluid interface near the rod, with this effect being more pronounced at lower elasticity numbers due to stronger inertial forces. Over time, the interface height begins to rise, driven by time-dependent viscoelastic effects. Once the climbing height reaches its maximum, it decreases and eventually stabilizes at a steady state.

To better understand this non-monotonic behavior, figure~\hyperref[inertia_steady_plot]{\ref{inertia_climbing_plot}($b$)} separately illustrates the inertial and viscoelastic contributions to the climbing height. Initially, the climbing height is negative due to the impulsive inertial effects at the onset. Over time, the inertial contribution continues to push the interface downward, while the viscoelastic contribution pushes it upward. At first, the viscoelastic effect dominates, causing the climbing height to increase. However, the viscoelastic contribution stabilizes more quickly than the inertial contribution. As a result, once the climbing height reaches its peak, it begins to decrease, driven by the slower approach of the inertial effects to their steady-state value. 

\begin{figure}
\hspace*{-0.27cm} 
\includegraphics[scale=0.303]{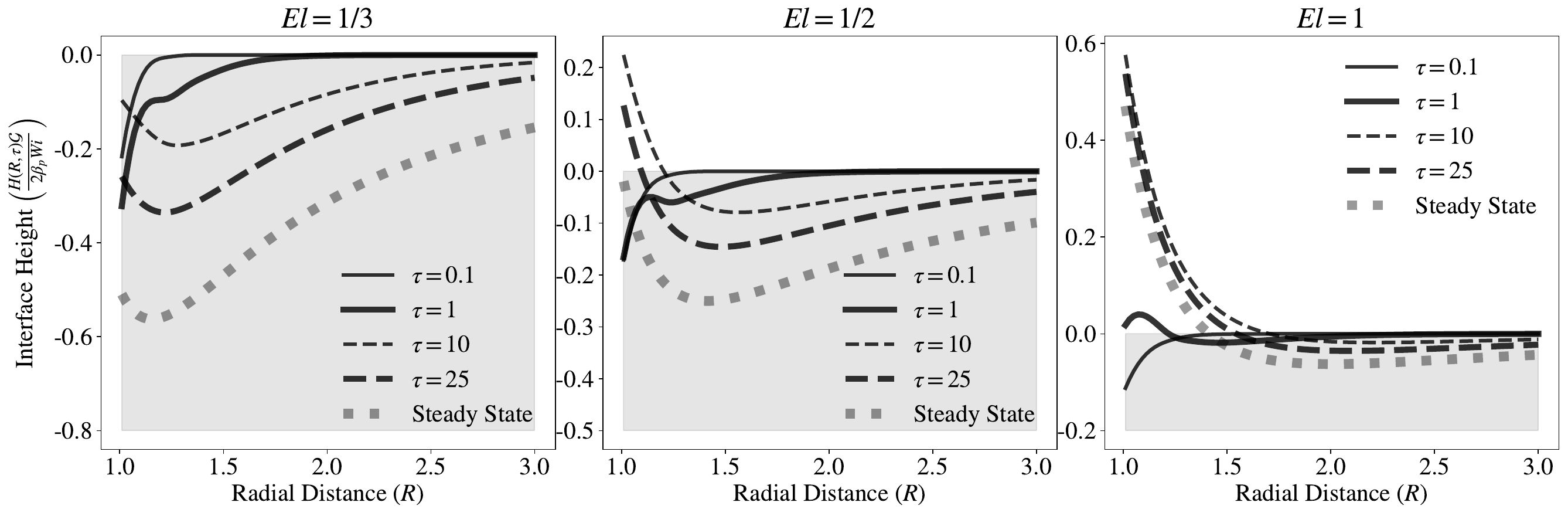}
\caption{Time evolution of the transient interface profiles around an infinitely long rotating rod in an Oldroyd-B fluid, incorporating small but finite inertial effects, shown at times $\tau = 0.1, 1, 10$ and $25$. We set $\beta_p = 0.5$ and explore three different values of the elasticity number $El= 1, 1/2$ and $1/3$.}
\label{inertia_transient_plot}
\end{figure}

To gain a clearer understanding of the transient interface profile, we present in figure~\ref{inertia_transient_plot}($a-c$) the time evolution of the free-surface profile (scaled by $2\beta_p Wi/\mathcal{G}$) for different elasticity numbers. Similar to our observations for the climbing height along the rod at $R = 1$ (figure~\hyperref[inertia_steady_plot]{\ref{inertia_climbing_plot}($a$)}), the initial rise in interface height after a short time is primarily driven by viscoelastic effects. In contrast, the subsequent decrease in height as the system approaches a steady state is predominantly governed by inertial effects. Depending on the spatial location of the interface, the viscoelastic fluid rises at varying rates due to the interplay between inertial and viscoelastic contributions. For $El = \frac{1}{3}, \frac{1}{2},$ and $1$, we observe that the interface profile exhibits non-monotonic behavior primarily within the range $ 1 \leq R \leq 1.5 $. At larger values of $R$, the interface deformation becomes smaller and more regular.

\section{Conclusions}\label{conclusions}
In this study, we explore the transient dynamics of the rod-climbing phenomenon, deriving the transient interface shape in the small-deformation regime for an Oldroyd-B fluid. Our analysis emphasizes the interplay between viscoelasticity and gravity, characterized respectively by the Weissenberg number ($Wi$) and the dimensionless gravity parameter ($\mathcal{G}$), and we find that the small-deformation scenario corresponds to cases where $Wi \ll 1$ and $\mathcal{G} \gg 1/Wi$. These conditions enable us to employ a perturbation expansion for the velocity field and the rise height in terms of $Wi$ with the aid of the domain perturbation method. In contrast, earlier work by~\citet{joseph1973} developed expansions based on the (small) angular rotation speed $\Omega$; see also \citep{more2023rod}. From a theoretical point of view, this approach is less desirable since $\Omega$ is a dimensional parameter, which complicates the distinction between limits arising from other variables, such as gravity and inertia.

Most experiments on the rod-climbing phenomenon focus on regimes where fluids rise higher than predicted in this study and form a thin film near the rod, deviating from the small-deformation scenario. One potential research direction to better understand this behavior is to analyze the phenomenon in the high-Weissenberg-number limit ($Wi \gg 1$), as significant climbing heights are observed at high angular velocities. Recent studies \citep{Hinch_Boyko_Stone_2024, Boyko_Hinch_Stone_2024} have investigated Oldroyd-B fluid flows in narrow, slowly varying contractions under similar conditions of $Wi \gg 1$, using the ultra-dilute limit and curvilinear transformations to simplify the equations. However, extending these methods to the rod-climbing problem may present a challenge as the flow is not unidirectional. Another possible research direction is to consider weak viscoelasticity ($Wi\ll1$) while incorporating moderate gravity effects ($\mathcal{G} < 1/Wi$), potentially revealing another distinguished limit where analytical solutions may be possible using asymptotic methods.

The methods developed in $\mathsection$~\ref{transient_section} to analyze the boundary layer in time and in $\mathsection$~\ref{with_inertia_section} to account for finite inertial effects are broadly applicable and not limited to the rod-climbing phenomenon. We anticipate that our approach for the Oldroyd-B model, which is also applicable to the Oldroyd-A model, will be valuable for studying a wider range of time-dependent problems, such as start-up flows in complex fluids. While most theoretical work on viscoelastic flows has focused on steady-state scenarios, our study aims to provide researchers with a versatile set of tools to gain deeper insight into how suspended polymers, and other similar microstructural elements, respond to unsteady flow fields. A key direction for future research is to extend our work to more complex constitutive equations that address the finite extensibility issues of the Oldroyd-B model, such as the Giesekus, FENE-P, FENE-CR, and PTT models. Expanding our results to these models will help improve the alignment with experimental data and enable more accurate rheological measurements using the rod-climbing phenomenon.
\backsection[Acknowledgements]{We thank Ehud Yariv for helpful discussions.}
\backsection[Funding]{
E.B. acknowledges the support by Grant No. 2022688 from the U.S.-Israel Binational Science Foundation (BSF). H.A.S. acknowledges the support from Grant No. CBET-2246791 from the U.S. National Science Foundation (NSF).}
\backsection[Declaration of interests]{The authors report no conflict of interest.}
\appendix

\section{Kernel of a linear operator $\mathcal{L}$}\label{kernel_of_L}
Here we prove a proposition about the kernel of the linear operator $\mathcal{L}$ used in $\mathsection$~\ref{transient_leading_section}. 
Suppose $\mathcal{L}$ be a linear operator defined by
\begin{equation}\label{operator_appendix}
\mathcal{L}(\textbf{F}(\textbf{X},t)) =  \beta_s \textbf{F}(\textbf{X},t) + \beta_p e^{-t} \int_{0}^{t} \textbf{F}(\textbf{X},s)e^{s}ds.
\end{equation}
If  $\mathcal{L}(\textbf{F}(\textbf{X},t)) = \textbf{0}$ for all $(\textbf{X},t) \in \mathbb{R}^3 \times [0,\infty)$ then $\textbf{F}(\textbf{X},t) = \textbf{0}$ for all $(\textbf{X},t) \in \mathbb{R}^3 \times [0,\infty)$.

The proof is as follows. Suppose that $\mathcal{L}(\textbf{F}(\textbf{X},t)) = \textbf{0}$ for all $\textbf{X} \in \mathbb{R}^3$ and $t \geq 0$. Evaluating the equation at $t = 0$, the second term on the right-hand-side of (\ref{operator_appendix}) vanishes, leaving us with $\textbf{F}(\textbf{X},0) = \textbf{0}$ as $\beta_s \neq 0$. We may now take a time derivative of the equation $e^{t}\mathcal{L}(\textbf{F}(\textbf{X},t)) = \textbf{0}$  to obtain
\begin{align}
\frac{d}{dt} \left(e^{t}\mathcal{L}(\textbf{F}(\textbf{X},t))\right) &= \frac{d}{dt} \left( \beta_s \textbf{F}(\textbf{X},t)e^{t} + \beta_p \int_{0}^{t} \textbf{F}(\textbf{X},s)e^{s}ds\right) \notag \\
&= \beta_s \frac{d}{dt}\left(\textbf{F}(\textbf{X},t)e^{t}\right) + \beta_p \textbf{F}(\textbf{X},t)e^{t} = \textbf{0}.
\end{align}
It follows that this first-order ordinary differential equation in terms of $\textbf{F}(\textbf{X},t)e^{t}$ yields the solution of the form,
\begin{equation}
\textbf{F}(\textbf{X}, t)e^{t} = \textbf{F}(\textbf{X}, 0)e^{-\frac{\beta_p}{\beta_s} t} = \textbf{0} \quad \Longrightarrow \quad \textbf{F}(\textbf{X},t) = \textbf{0} \quad \text{for all} \quad \textbf{X} \in \mathbb{R}^3 \text{ and } t \in [0,\infty).
\end{equation}

\section{Evaluating the hoop stress}\label{hoop_stress_calculation}
Here, we provide detailed calculations for the expression of $\nabla \cdot \mathsfbi{A}^{(2)}$ in (\ref{inertia_transient_polymeric_force}). We begin by noting that the leading-order velocity field takes the form $\textbf{U}^{(0)} = U^{(0)}(R, \tau)\textbf{e}_{\theta}$. This enables us to write down the rate-of-strain tensor associated with the leading-order velocity field,
\begin{equation}
    \boldsymbol{\mathit{E}}^{(0)} = \frac{1}{2}\left(\frac{\partial U^{(0)}}{\partial R} - \frac{U^{(0)}}{R}\right)\left(\textbf{e}_{r}\textbf{e}_{\theta} + \textbf{e}_{\theta}\textbf{e}_{r}\right) = \frac{R}{2}\frac{\partial}{\partial R}\left(\frac{U^{(0)}}{R}\right)\left(\textbf{e}_{r}\textbf{e}_{\theta} + \textbf{e}_{\theta}\textbf{e}_{r}\right).
\end{equation}
Using (\ref{first_order_transient_conformation}), we express the first-order correction to the conformation tensor as
\begin{equation}
    \mathsfbi{A}^{(1)} = \left(e^{-\tau}R\int_{0}^{\tau}\frac{\partial}{\partial R}\left(\frac{U^{(0)}}{R}\right)e^{s}\:ds\right)\left(\textbf{e}_{r}\textbf{e}_{\theta} + \textbf{e}_{\theta}\textbf{e}_{r}\right).
\end{equation}
At the second order, the evolution of the conformation tensor (\ref{conformation_stretched}) is
\begin{align}\label{second_order_conformation_transient_appendix}
\mathsfbi{A}^{(2)} + \frac{\partial \mathsfbi{A}^{(2)}}{\partial \tau} =  -\textbf{U}^{(0)}\cdot \nabla \mathsfbi{A}^{(1)} + \mathsfbi{A}^{(1)} \cdot \nabla\textbf{U}^{(0)}+(\nabla \textbf{U}^{(0)})^{\rm T}\cdot \mathsfbi{A}^{(1)} + 2\boldsymbol{\mathit{E}}^{(1)}. 
\end{align}
Next, we calculate each term on the right-hand side of (\ref{second_order_conformation_transient_appendix}).
Since $\textbf{U}^{(0)}$  is azimuthal, we derive
\begin{equation}
    \textbf{U}^{(0)}\cdot \nabla \mathsfbi{A}^{(1)} = \frac{U^{(0)}}{R} \frac{\partial \mathsfbi{A}^{(1)}}{\partial \theta} = \left(2e^{-\tau}U^{(0)}\int_{0}^{\tau}\frac{\partial}{\partial R}\left(\frac{U^{(0)}}{R}\right)e^{s}\:ds\right)\left(\textbf{e}_{\theta}\textbf{e}_{\theta} - \textbf{e}_{r}\textbf{e}_{r}\right).
\end{equation}
On the other hand, 
\begin{equation}
    \mathsfbi{A}^{(1)} \cdot \nabla\textbf{U}^{(0)}+(\nabla \textbf{U}^{(0)})^{\rm T}\cdot \mathsfbi{A}^{(1)} = 2e^{-\tau}\left(\int_{0}^{\tau}\frac{\partial}{\partial R}\left(\frac{U^{(0)}}{R}\right)e^{s}\:ds\right)\left(R\frac{\partial U^{(0)}}{\partial R}\textbf{e}_{\theta}\textbf{e}_{\theta} - U^{(0)}\textbf{e}_{r}\textbf{e}_{r}\right).
\end{equation}
Thus, (\ref{second_order_conformation_transient_appendix}) can now be rewritten as,
\begin{equation}\label{second_order_conformation_transient_appendix_simp}
    \mathsfbi{A}^{(2)} + \frac{\partial \mathsfbi{A}^{(2)}}{\partial \tau}  = 2\boldsymbol{\mathit{E}}^{(1)} + 2e^{-\tau}R^2\frac{\partial}{\partial R}\left(\frac{U^{(0)}}{R}\right)\left(\int_{0}^{\tau}\frac{\partial}{\partial R}\left(\frac{U^{(0)}}{R}\right)e^{s}\:ds\right) \textbf{e}_{\theta}\textbf{e}_{\theta}.
\end{equation}
Multiplying both sides of (\ref{second_order_conformation_transient_appendix_simp}) with $e^{\tau}$ and then integrating with respect to $\tau$, we arrive at
\begin{equation}
    \mathsfbi{A}^{(2)} = 2e^{-\tau}\int_{0}^{\tau}\boldsymbol{\mathit{E}}^{(1)} e^{s}\: ds+ 2e^{-\tau}\int_{0}^{\tau}R^2\frac{\partial}{\partial R}\left(\frac{U^{(0)}}{R}\right)\left(\int_{0}^{S}\frac{\partial}{\partial R}\left(\frac{U^{(0)}}{R}\right)e^{s}\:ds\right)\: dS \:  \textbf{e}_{\theta}\textbf{e}_{\theta}.
\end{equation}
Thus, we conclude that 
\begin{equation}
    \nabla \cdot \mathsfbi{A}^{(2)} = e^{-\tau}\int_{0}^{\tau}\nabla^2 \textbf{U}^{(1)} e^{s}\: ds- 2e^{-\tau}\int_{0}^{\tau}R\frac{\partial}{\partial R}\left(\frac{U^{(0)}}{R}\right)\left(\int_{0}^{S}\frac{\partial}{\partial R}\left(\frac{U^{(0)}}{R}\right)e^{s}\:ds\right)\: dS \:  \textbf{e}_{r}.
\end{equation}
\bibliographystyle{jfm}
\bibliography{literature}

\end{document}